\documentclass{aa}  
\usepackage{geometry}
\usepackage{float}
\usepackage{titlesec}
\usepackage{amssymb,amsmath}
\usepackage[]{graphicx}
\usepackage{caption}
\usepackage{subcaption}
\usepackage[]{xcolor}
\usepackage{booktabs, multirow}
\usepackage{graphicx}
\usepackage{subcaption}
\usepackage{txfonts}
\usepackage{adjustbox}
\usepackage{amsmath}
\usepackage[colorlinks,linkcolor=black,urlcolor=gray,citecolor=blue]{hyperref}
\usepackage{natbib}
\begin{document}

   \title{The head-tail radio galaxy and revived fossil plasma in Abell 1775}


   \author{A. Bushi
          \inst{\ref{11}}\fnmsep\inst{\ref{22}}
          \and
          A. Botteon\inst{\ref{22}}
          \and
          D. Dallacasa\inst{\ref{11}}\fnmsep\inst{\ref{22}}
          \and
          R. J. van Weeren\inst{\ref{33}}
          \and
          T. Venturi \inst{\ref{22}}
          \and
              M. Br\"uggen\inst{\ref{44}}
          \and
          F. Gastaldello\inst{\ref{55}}
          \and
          S. Giacintucci\inst{\ref{66}}
          }

   \institute{Dipartimento di Fisica e Astronomia (DIFA), Università di Bologna, via Gobetti 93/2, 40129 Bologna, Italy \label{11}\\
              \email{ardiana.bushi@gmail.com}
         \and
             Istituto Nazionale di Astrofisica (INAF) – Istituto di Radioastronomia (IRA), via Gobetti 101, 40129 Bologna, Italy \label{22}
        \and
            Leiden Observatory, Leiden University, PO Box 9513, NL-2300 RA Leiden, The Netherlands \label{33}
        \and
           Hamburger Sternwarte, Universität Hamburg, Gojenbergsweg 112, D-21029 Hamburg, Germany \label{44}
        \and
           INAF - IASF Milano, via A. Corti 12, I-20133 Milano, Italy \label{55}
        \and
           Naval Research Laboratory, 4555 Overlook Avenue SW, Code 7213, Washington, DC 20375, USA \label{66}
            }

   \date{Received XXX; accepted YYY}

 
  \abstract
   {Head-tail radio galaxies are characterized by a head, corresponding to an elliptical galaxy, and two radio jets sweeping back from the head, forming an extended structure behind the host galaxy that is moving through the intracluster medium (ICM). This morphology arises from the interaction between the diffuse radio-emitting plasma and the surrounding environment. Sometimes, in galaxy clusters, revived fossil plasma can be found, and it traces old active galactic nucleus ejecta with a very steep spectrum that has been re-energized through processes in the ICM, unrelated to the progenitor radio galaxy.}
   {We focus on the study of the central region of Abell 1775, a galaxy cluster in an unclear dynamical state at a redshift of z = 0.07203. It hosts two giant radio-loud elliptical galaxies, the head-tail radio galaxy that ‘breaks’ at the position of a cold front detected in the X-rays, filamentary revived fossil plasma, and central diffuse emission.
This study aims to investigate and constrain the spectral properties and trends along the head-tail, as well as the revived fossil plasma, to better understand the formation process of the non-thermal phenomena in A1775.}
   {We make use of observations at three frequencies performed with LOFAR at 144 MHz, and new deep uGMRT observations at 400 MHz and at 650 MHz.}
   {We observe an overall steepening along the tail of the head-tail radio galaxy. In the radio colour-colour diagram, ageing models reproduce the emission of the head-tail. An unexpected brightness increase at the head of the tail suggests a complex bending of the jets.
We derived the equipartition magnetic field and minimum pressure along the tail.
We recovered the structure of the revived fossil plasma, which appears as thin filaments with ultra-steep spectra.}
   {We show that high-sensitivity, high-resolution observations at low frequencies are essential for detecting the full extent of the tail, enabling a deeper spectral analysis and resolving the structure and spectral properties of revived fossil plasma.}


   \maketitle
%

\section{Introduction}

In galaxy clusters, radio observations reveal synchrotron emission originating from both individual radio galaxies and diffuse radio sources. The radio emission reveals the presence of relativistic cosmic ray electrons (CRe) and magnetic fields in the hot ionized ($10^{7}-10^{8} \ \text{K} $) and tenous ($10^{-3} - 10^{-4} \ \text{cm}^{-3}$) gas of the intracluster medium (ICM). 

Some radio galaxies in clusters are extended (up to the megaparsec scale), and often show signs of interaction with the ICM, resulting in wide-angle tail, narrow-angle tail, and head-tail morphologies depending on the bending of the jets \citep{Miley_1980}. The head-tail radio galaxies discovered by \cite{Ryle_Windram_1968} are characterized by a head, identified with the optical galaxy, and two tails that consist of the relativistic plasma deposited by the jets emanating from the active galactic nucleus (AGN). These jets are swept back from the parent galaxy and often merge. The bending of the jets or tails is due to the radio-emitting plasma being deflected by the ICM via ram pressure while the host galaxy is moving through the cluster \citep[e.g.,][]{RudnickOwen1976,OwenRudnick1976,OdeaOwen1985,  Dehghan_2014, Sebastian2017, Terni_de_Gregory_2017, Missaglia_2019, Odea2023}.

The spectral index of head-tail radio galaxies steepens from the head towards the tail, indicating that the oldest relativistic plasma is situated furthest from the host galaxy. The CRe in the tail lose energy through synchrotron and inverse-Compton emission. Low-frequency observations are crucial for tracing low-energy CRe, which have longer cooling times. With the advent of sensitive observations at low frequency ($<$1 GHz), there are now more detailed studies of the most extended tails, which reach up to several hundred kiloparsecs (kpc) \citep[e.g.,][]{Sebastian2017,deGasperin_2017,Wilber_2018,Srivastava_2020,Botteon_2021,Edler_2022, Lusseti_2024, Bruno2024, Koribalski2024}.     

The ICM dynamics plays an important role in the morphology and re-acceleration of relativistic electrons of tailed cluster radio galaxies. These electrons can gain energy through a variety of processes, such as turbulent re-acceleration \citep[e.g.,][]{Loken_1995}, adiabatic compression \citep[e.g.,][]{Enblin_and_Gobal-Krishna_2001,EnblinBrugen2002}, shock re-acceleration \citep[e.g.,][]{vanWeeren2017}, or mechanisms that can barely balance electron radiative losses caused by turbulence in the ICM  within the radio tails \citep[e.g.,][]{deGasperin_2017}. Recent radio and X-ray observations provide clear evidence of shocks or cold fronts co-located with radio galaxy tails in many clusters, explaining the disturbed morphology of the radio emission \citep{Chibueze2021,Botteon_2021,Giacintucci2022,Brienza2022,Lee2023,Bruno2024,Vazza_2024, Koribalski2024}.                  

Revived fossil plasma is a class of diffuse radio source that can also be found in galaxy clusters, where its emission traces radio plasma of AGN origin that has somehow been re-energized. Re-energization is possible through processes such as in situ re-acceleration of fossil plasma and cosmic rays (CR) in the ICM, unrelated to the nuclear activity of the radio galaxy. These sources exhibit a very steep spectrum and their resolved spectral properties do not show any clear trends \citep{vanWerren2011,Cohen_and_Clarke_2011,Kale_and_Dwarakanath_2012,Mandal_2020}. Both radio sources, the head-tail radio galaxies, and revived fossil plasma have steep spectra that are best studied through low-frequency observations.

Abell 1775 is a galaxy cluster classified with a richness of R=2. It is located at $z$ = 0.072 and hosts two distinctive radio galaxies at its centre.
The cluster belongs to the Boötes supercluster \citep{Einasto_1997,Chow-Martinez_2014} and was initially identified as containing two interacting subclusters through optical studies based on the radial velocities of the 51 member galaxies \citep{Oegerle_1995,Kopylova_2009}. The radial velocity difference between the subclusters was found to be $\sim2900 \ \text{km} \ \text{s}^{-1}$.
Subsequent studies, with larger samples of cluster members and combined photometric data from the Sloan Digital Sky Survey (SDSS) and the Beijing-Arizona-Taiwan-Connecticut (BATC) \citep{Zhang_2011}, revealed the possibility of a ternary system. 
A1775 has an X-ray luminosity of $1.6 \times 10^{44} \ \text{erg} \ \text{s}^{-1}$ and a flux of $1.25 \times 10^{-11}$  $\text{erg} \ \text{s}^{-1} \ \text{cm}^{-2}$ in the $0.1 - 2.4$ keV band \citep{Ebeling_1998}. In the second Planck catalogue of Sunyaev-Zeldovich (SZ) sources \citep{Planck_Collaboration_2016}, it is reported with the name PSZ2G031.93+78.71 and the mass $M_{500}\footnote{$M_{500}$ represents the mass contained within the radius, where the mean density is 500 times the critical density of the universe.} = (2.72 \pm 0.24) \times 10^{14} \ \text{M}_{\odot}$, and it is also part of CHEX-MATE\footnote{\url{http://xmm-heritage.oas.inaf.it/}} cluster sample \citep{CHEX2021}.

The head-tail radio galaxy of Abell 1775 has been studied by \cite{Owen_Ledlow_1997,Giovannini_and_ferretti_2000,Giacintucci_2007, Terni_de_Gregory_2017}. The central region of this cluster was previously studied by \cite{Botteon_2021} using radio observations of the LOw Frequency ARray (LOFAR) at 144 MHz, the Giant Metrewave Radio Telescope (GMRT) at 235 and 610 MHz, and the Very Large Array (VLA) at 1.4 GHz, along with X-ray data from Chandra. They revealed the presence of revived fossil plasma, an 800 kpc head-tail radio galaxy (two times longer than what was reported by \cite{Giacintucci_2007}), and central diffuse emission. Two giant elliptical galaxies are the most prominent cluster members, with a radial-velocity difference of $\sim1800 \ \text{km} \ \text{s}^{-1}$. Both galaxies are radio-loud, with B1339+266B being the host of the head-tail radio galaxy.
With Chandra X-ray observations, \cite{Botteon_2021} found evidence of gas motions in the cluster centre and highlighted the presence of a cool, spiral-like structure in the ICM induced by the gas motions. The emission of the radio tail has a sudden change in morphology roughly at the location of a cold front detected in the X-rays. This study suggests an interaction between the tail and the surrounding medium. The radio features mentioned and the cold front in the X-ray in the central region of Abell 1775 are shown in Figure \ref{fig:Head-tail-Botteon}. The overall results such as the detection of the arc-shaped cold front from X-ray observations by \cite{Botteon_2021} are consistent with a recent study by \cite{Hu_2021} done by analyzing deep Chandra and XMM-Newton archive data and by using hydrodynamic simulations. The simulations found that after the first pericentric passage between two clusters, a merger mass ratio of 5 best matches the observed X-ray emission and temperature distributions. In this paper, we have made use of observations taken with LOFAR at 144 MHz along with new high-sensitivity and high-angular-resolution uGMRT observations at 400 MHz and 650 MHz to study the region of the tail farthest from the AGN and the revived fossil plasma, which were clearly detected only with LOFAR in the previous work by \cite{Botteon_2021}.

The structure of this paper is as follows. In Section \ref{2}, we outline the radio telescope observations of LOFAR and uGMRT, as well as the data reduction and imaging procedures. In Section \ref{3}, we introduce the final radio images at high and low resolution for the central region of Abell 1775 and report the analysis for the head-tail and revived fossil plasma. In Section \ref{4}, we present a discussion of our findings.
Section \ref{5} summarizes the main results and the salient conclusions of this study.

We adopted a $\Lambda CDM$ cosmology with $\Omega_{\Lambda} = 0.7$, $\Omega_m = 0.3$, and $H_0 = 70 \ \text{km} \ \text{s}^{-1} \ \text{Mpc}^{-1}$, in which 1 arcsec corresponds to 1.372 kpc at $z = 0.072$. The spectral index is defined by the convention $S_\nu \propto \nu^{-\alpha}$ for radio synchrotron emission, with $\alpha > 0$.


\section{Observations and data reduction \label{2}}
We used LOFAR observations centred at 144 MHz, uGMRT band 3 at 400 MHz, and uGMRT band 4 at 650 MHz (see Tab. \ref{tab:Obs} for a summary). In the upcoming sections, we briefly describe the data reduction procedure.

\begin{table*} [h!]
   \captionsetup{labelfont=bf}  
   \caption{Observations.}
   \label{tab:Obs}
   \begin{center} 
     \begin{tabular}{l l l c c c} 
       \hline
       Telescope & Observation Date & Time [h] & \multicolumn{1}{c}{Frequency} & \multicolumn{1}{c}{Central Frequency} & \multicolumn{1}{c}{PI} \\
                 &           &          & \multicolumn{1}{c}{[MHz]} & \multicolumn{1}{c}{[MHz]} \\
       \hline
       LOFAR (P207+25) & $2017$ Oct. $11$ & 8 & 120-168 & 144 & T. Shimwell\\
       LOFAR (P204+25) & $2019$ Apr. $04$ & 8 & 120-168 & 144 & T. Shimwell\\
       LOFAR (P207+27) & $2017$ Feb. $09$ & 8 & 120-168 & 144 & T. Shimwell\\
       LOFAR (P204+27) & $2020$ July $30$ \& $2020$ Aug. $02$ & 4+4 & 120-168 & 144 & T. Shimwell\\
       uGMRT band 3 & $2021$ May $25$ & 6 & 300-500 & 400 & A. Botteon\\
       uGMRT band 4 & $2021$ May $30$ & 6 & 550-750 & 650 & A. Botteon\\
       \hline
     \end{tabular}
   \end{center}
\end{table*}

\subsection{LOFAR data}
A1775 is located within $2.5^{\circ}$ from the centre of the four pointings (P207+25, P204+25, P207+27, and P204+27) of the LOw Frequency ARray (LOFAR) Two-meter Sky Survey \citep[LoTSS;][]{Shimwell_2017, Shimwell_2019, Shimwell_2022}. LoTSS is a deep imaging low-frequency radio survey of the northern sky with a nominal sensitivity of 0.1 mJy $\text{beam}^{-1}$ at a resolution of $6\arcsec$. Each pointing was observed using LOFAR high band antennas operating at 120 - 168 MHz with an integration time of eight hours, book-ended by 10-minute flux calibrator scans.
We used the data from the Second LoTSS Data Release  \citep[LoTSS-DR2;][]{Shimwell_2022}, which has undergone additional improvements in terms of image fidelity, particularly for faint diffuse structures. Abell 1775 belongs to the second Planck catalogue of SZ sources \citep[PSZ2;][]{Planck_Collaboration_2016} and is part of the LoTSS-DR2/PSZ2 sample which was presented in \cite{Botteon_2022}. In this work, we use the same calibrated data as is presented in \cite{Botteon_2021} and \cite{Botteon_2022}. For clarity, below we briefly report the main steps of the data analysis.

The processing of the LOFAR data for each pointing observation was done using the reduction pipeline v2.2  \citep[see][]{Shimwell_2022,Tasse_2021}, which was developed by the LOFAR Surveys Key Science Project team (\textsc{prefactor}, \citealt{vanWeeren_2016,Williams_2016,deGasperin2019}; \textsc{killMS} \citealt{Tasse_2014a,Tasse2014b,SmirnovTasse2015}; \textsc{DDFacet} \citealt{Tasse_2018}) to correct for both direction-independent and direction-dependent effects. To improve the calibration towards the target, a second step in the data reduction was performed, subtracting the sources outside a $33\arcmin \times 33\arcmin$ region that contains the target (A1775) from the visibility data with the method described in \cite{vanWeeren_2021}.

With the already calibrated data, we jointly imaged the four observations and deconvolved using \textsc{WSClean v3.1} \citep{Offringa_2014} with the deconvolution algorithm called ‘joined-channel deconvolution’. Final images at the central observing frequency of 144 MHz were obtained using the multi-scale and multi-frequency synthesis deconvolution scheme \citep{Offringa_and_Smirnov_2017} with Briggs weightings \citep{Briggs_1995} of \textsf{robust}=-1.5 for the high-resolution image ($6\arcsec \times 6\arcsec$). We applied the Gaussian \textsf{taper} and increased the \textsf{robust} parameter for low-resolution images ($15\arcsec \times 15\arcsec$). The final resolution for the images was obtained after the smoothing process and will be further discussed in Sec.\ref{2.3}.
During imaging, we made use of cleaning masks generated using the \textsc{breizorro}\footnote{\url{https://github.com/ratt-ru/breizorro} \label{brei}} tool. 

It is known that the LOFAR flux density scale can show systematic offsets \citep{Hardcastle_2016}. Therefore, in this analysis, a correction factor of 0.91 was applied to the LOFAR flux densities, as was established by \cite{Botteon_2022}. The systematic flux scale uncertainty assumed for LOFAR is $10\%$ \citep{Shimwell_2022}.

\subsection{uGMRT data}
Compared to the archival `legacy' GMRT data used by \cite{Botteon_2021}, we used higher-sensitivity and higher-angular-resolution observations carried out with the upgraded Giant Metrewave Radio Telescope (uGMRT). This instrument has upgraded receivers with wide-band capabilities that now cover frequencies from 120 to 1450 MHz \citep{Gupta_2017}. 

Abell 1775 was observed at 300-500 MHz (band 3) and 550-750 MHz (band 4), for 6 hours in each band. We used the \textsc{SPAM}\footnote{\url{https://www.intema.nl/doku.php?id=huibintema:spam:pipeline} \label{spam}} pipeline for the direction-dependent calibration of the data \citep{Intema_2014,Intema_2017}. As this pipeline was originally developed to process narrow-band data, we split the bandwidth of 200 MHz of the uGMRT datasets into six and four sub-bands for band 3 and 4, respectively. Each of these sub-bands was processed independently using \textsc{SPAM}.

Once we calibrated the data, we jointly deconvolved the six sub-bands of uGMRT band 3 data and created high- and low-resolution images at the central frequency at 400 MHz. Following a process similar to that applied to the LOFAR data, we employed the multi-scale, multi-frequency deconvolution scheme with Briggs weightings of \textsf{robust}=-1.5 for the high-resolution image of $6\arcsec \times 6\arcsec$ and Gaussian \textsf{taper} with \textsf{taper}=11 for the low-resolution image with a resolution at $15\arcsec \times 15\arcsec$. The same procedure was followed for uGMRT band 4 data, which had four frequency sub-bands, leading to images with a central frequency of 650 MHz. We deconvolved images at high and low resolution with the same \textsf{robust} and \textsf{taper} values as was employed for band 3. The final resolutions for images in each band were achieved following the steps outlined in the next section. As for the LOFAR imaging, cleaning masks generated with \textsc{breizorro} were used to guide the cleaning. The final images were corrected for the primary beam attenuation.

We calculated the spectra of a few selected point sources across three frequencies LOFAR at 144 MHz, uGMRT at 400 MHz, and VLA public archive data at 1.4 GHz from the FIRST survey \citep{First_1995}. We excluded the data points from uGMRT band 4 from the fit calculation because they consistently lie above the best-fit lines, indicating a positive offset in the uGMRT band 4 flux scale. To correct this, we determined a correction factor of 0.89, which we applied to the uGMRT band 4 flux density.
The systematic uncertainties due to residual amplitude errors were set to $10\%$ for band 3 and $5\%$ for band 4 \citep{Chandra_2004}. 

\subsection{Imaging \label{2.3}}
To ensure a reliable comparison between the flux densities measured in the LOFAR and uGMRT images, the $uv$ coverages of the instruments need to be comparable. LOFAR has a dense array with short baselines. We examined the $uv$ coverage of uGMRT data to determine its short baseline coverage. For all the images, we decided to use a common lower $uv$ cut of $80 \lambda$ (\textsf{minuv-l=80}).
To conduct further spectral analysis, several pre-processing steps were required, using the LOFAR image as a reference. These steps included image alignment to correct for any positional offsets, followed by re-gridding on the same pixel layout. Finally, we ensured that all images have the same restoring beam by convolving the images to a common resolution of 6" and 15". Table \ref{tab:Noise} shows a summary of the properties of the images that we used in the paper.

\begin{table}[h!]
  \centering
  \captionsetup{labelfont=bf}
  \caption{The resolutions and noise levels of the final images from LOFAR, uGMRT band 3 and 4 that are shown in Fig. \ref{HighandLowResolutionFinalImages}.}
  \label{tab:Noise}
  \begin{tabular}{
    |c|c|c|c|
  }
    \toprule
    \multicolumn{1}{|c|}{\text{Resolution}} & \multicolumn{1}{c|}{\text{Telescope}} & \multicolumn{1}{c|}{\text{$\sigma_{\text{rms}}$}} \\
\multicolumn{1}{|c|}{\text{[$\arcsec \times \arcsec$]}} & \multicolumn{1}{c|}{} & \multicolumn{1}{c|} {\text{[$\mu  \text{Jy} \ \text{beam}^{-1}$]}} \\
    \midrule

                    $6\times6$ & LOFAR & 164 \\
                       & uGMRT band 3 & 35\\
                       & uGMRT band 4 & 13\\
                    $15\times15$ & LOFAR & 149 \\
                       & uGMRT band 3  & 73  \\
                       & uGMRT band 4 & 34  \\
    \bottomrule
  \end{tabular}
\end{table}

\section{Results \label{3}}
\begin{figure}
    \captionsetup{labelfont=bf} 
    \centering
    \includegraphics[width=0.48\textwidth]{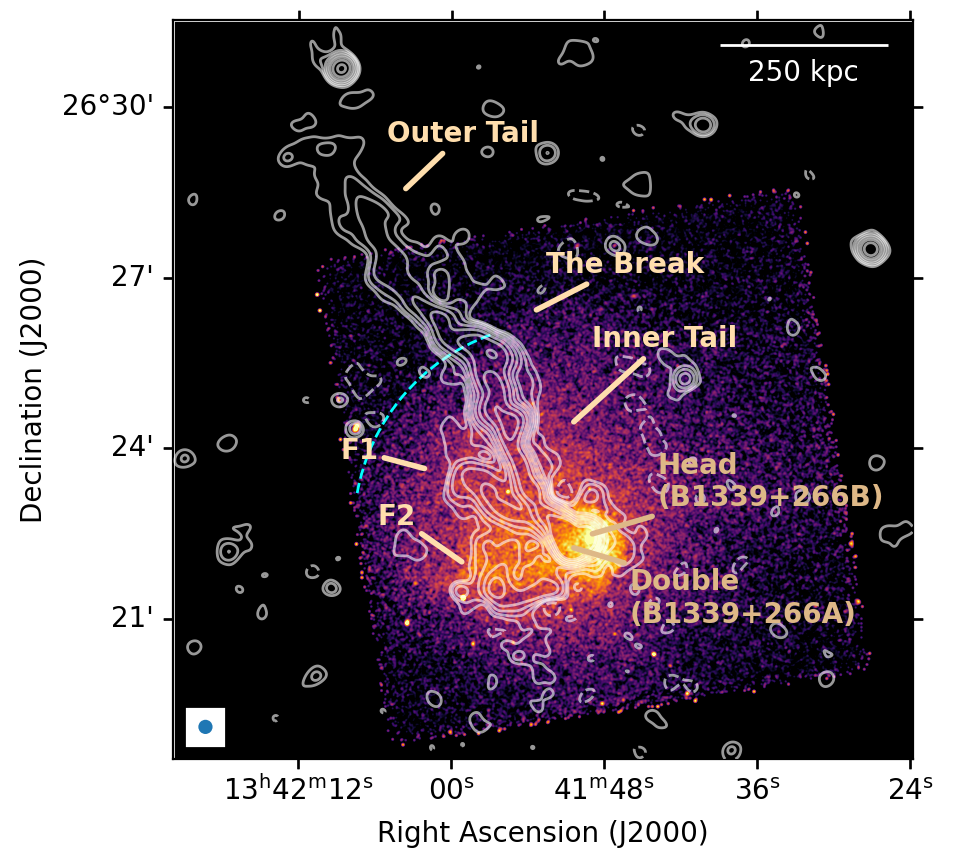}
    \caption{Low-resolution ($15\arcsec \times 15\arcsec$) uGMRT 400 MHz radio contours overlaid on the Chandra X-ray image in the 0.5–2.0 keV band \citep{Botteon_2021} of A1775, with main features labelled. White radio contours start from 3$\sigma$ and are spaced by a factor of 2. The beam is shown in the bottom left corner together with the scale bar at the top right.}
    \label{fig:Head-tail-Botteon}
\end{figure}

In Figure \ref{fig:Head-tail-Botteon}, we present the radio contours of Abell 1775's central region, featuring the components of the head-tail radio galaxy that we label as is done by \cite{Botteon_2021}: the inner tail, the break, and the outer tail. Additionally, the figure highlights the revived fossil plasma filaments (F1 and F2), and in the central region of this cluster two giant elliptical galaxies, one hosting the head of the long head-tail radio galaxy (B1339+266B, the Head) and the other a double radio source (B1339+266A, the Double). The two galaxies are separated by a projected distance of 32 kpc \cite{Parma_1991}. Radio contours in Fig. \ref{fig:Head-tail-Botteon} are overlaid on the Chandra X-ray image, where the cold front is marked with a dashed blue arc and traces the X-ray surface brightness discontinuity detected by \cite{Botteon_2021}.

\begin{figure*}[h!]
    \captionsetup{labelfont=bf} 
    \centering

    \begin{subfigure}{0.33\textwidth}
        \centering
        \includegraphics[width=\linewidth]{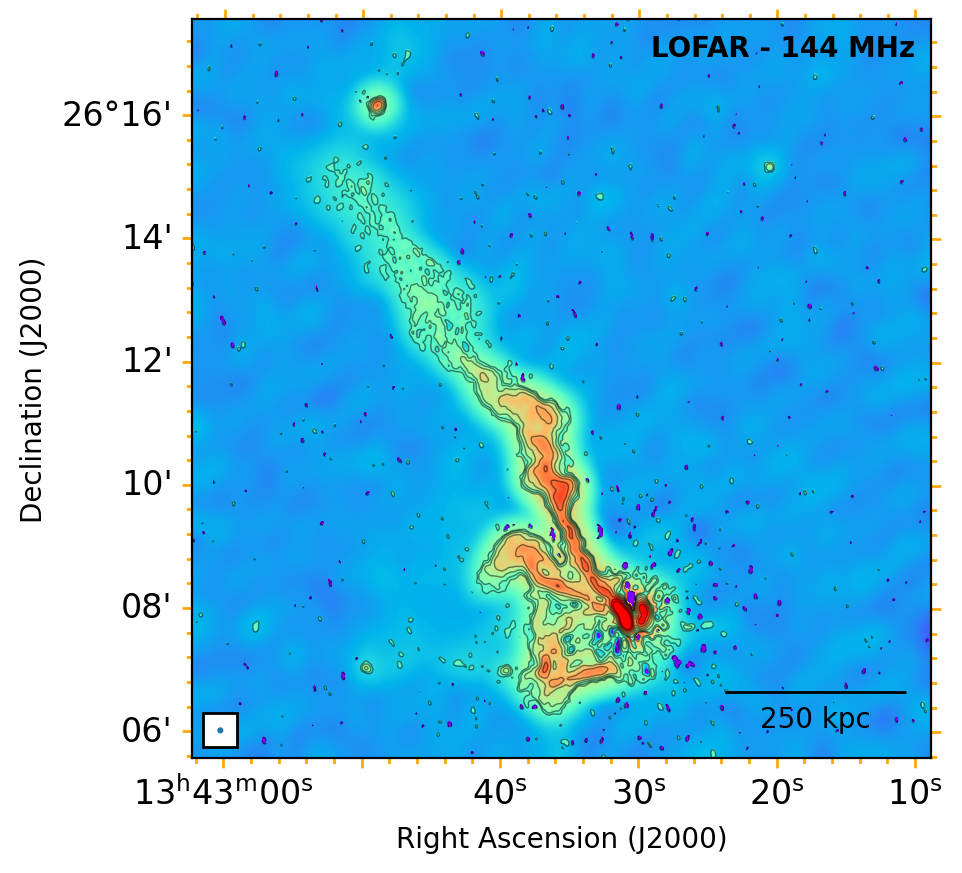}
    \end{subfigure}
    \hfill
    \begin{subfigure}{0.33\textwidth}
        \centering
        \includegraphics[width=\linewidth]{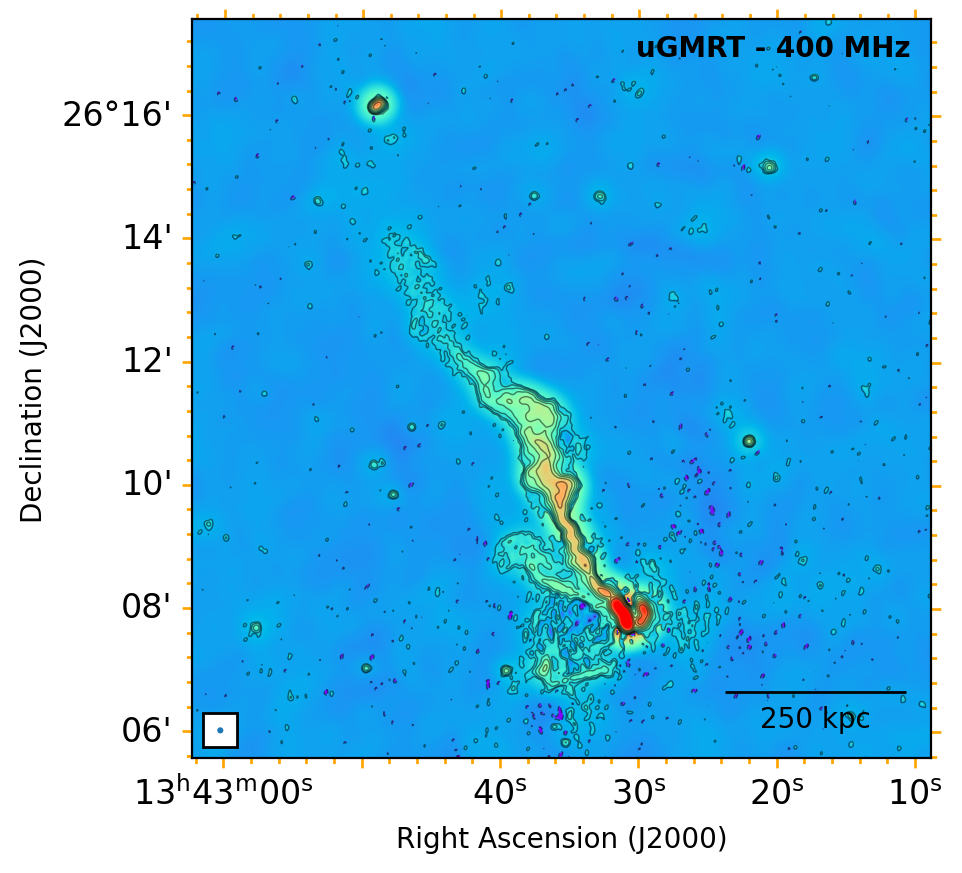}
    \end{subfigure}
    \hfill
    \begin{subfigure}{0.33\textwidth}
        \centering
        \includegraphics[width=\linewidth]{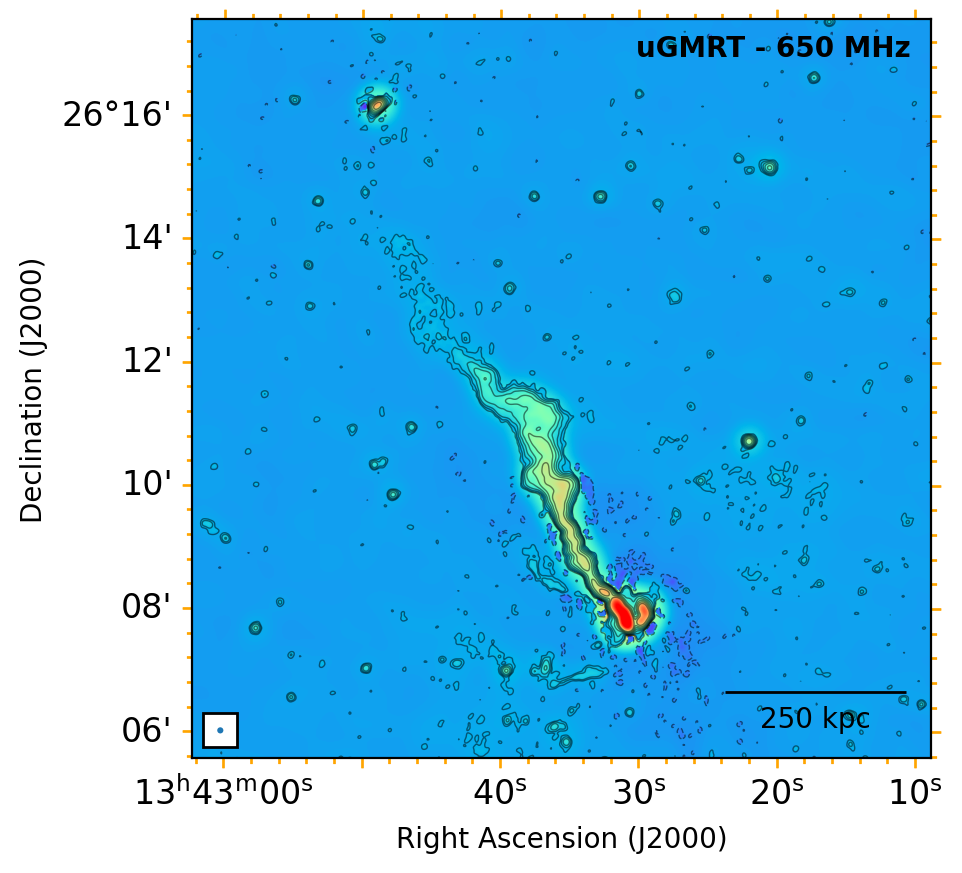}
    \end{subfigure}
    
    \vspace{0.5cm} 

    \begin{subfigure}{0.33\textwidth}
        \centering
        \includegraphics[width=\linewidth]{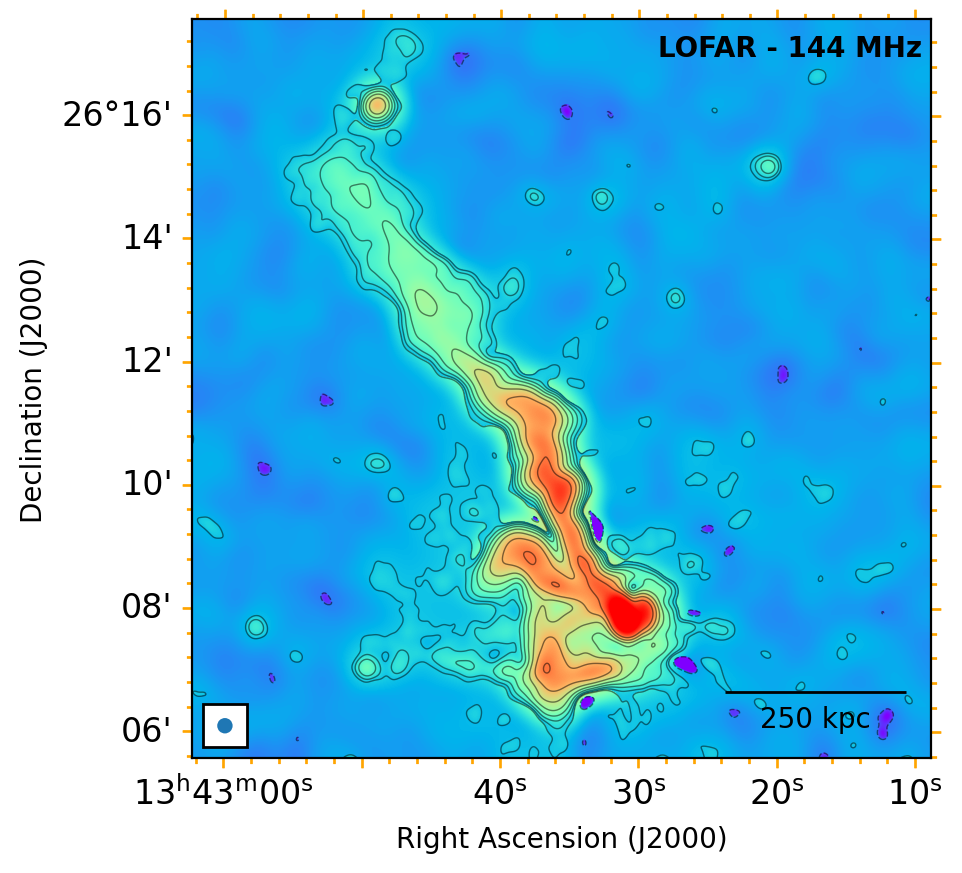}
    \end{subfigure}
    \hfill
    \begin{subfigure}{0.33\textwidth}
        \centering
        \includegraphics[width=\linewidth]{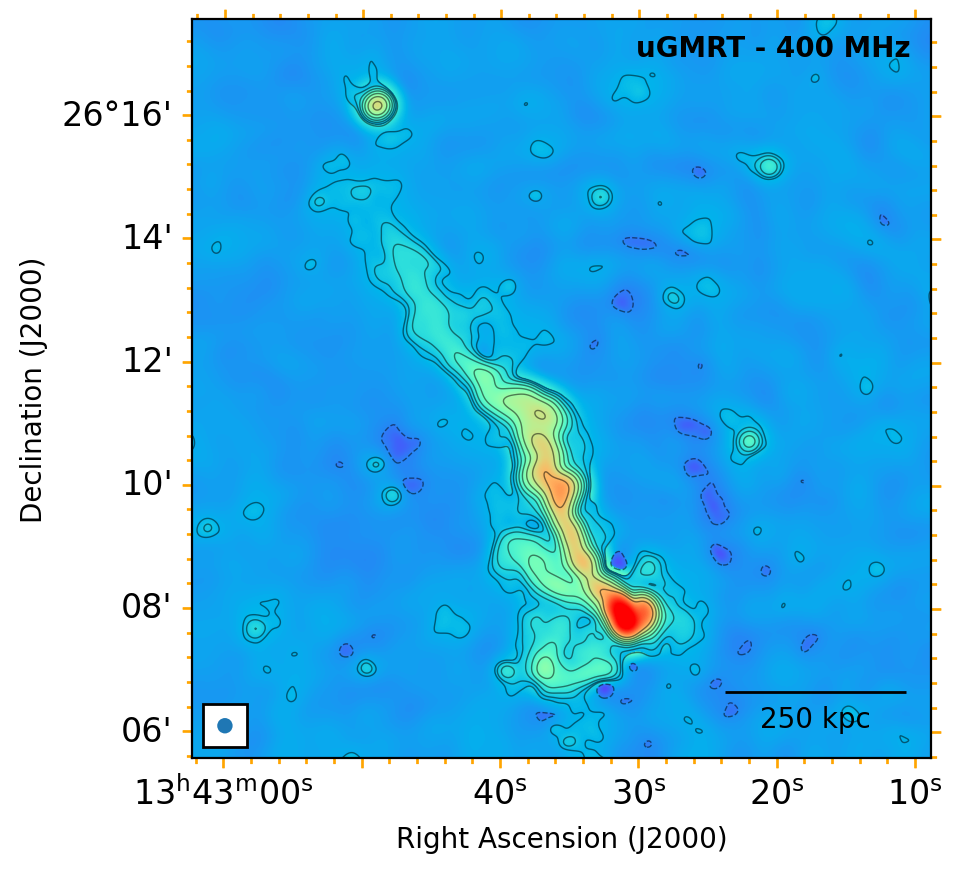}
    \end{subfigure}
    \hfill
    \begin{subfigure}{0.33\textwidth}
        \centering
        \includegraphics[width=\linewidth]{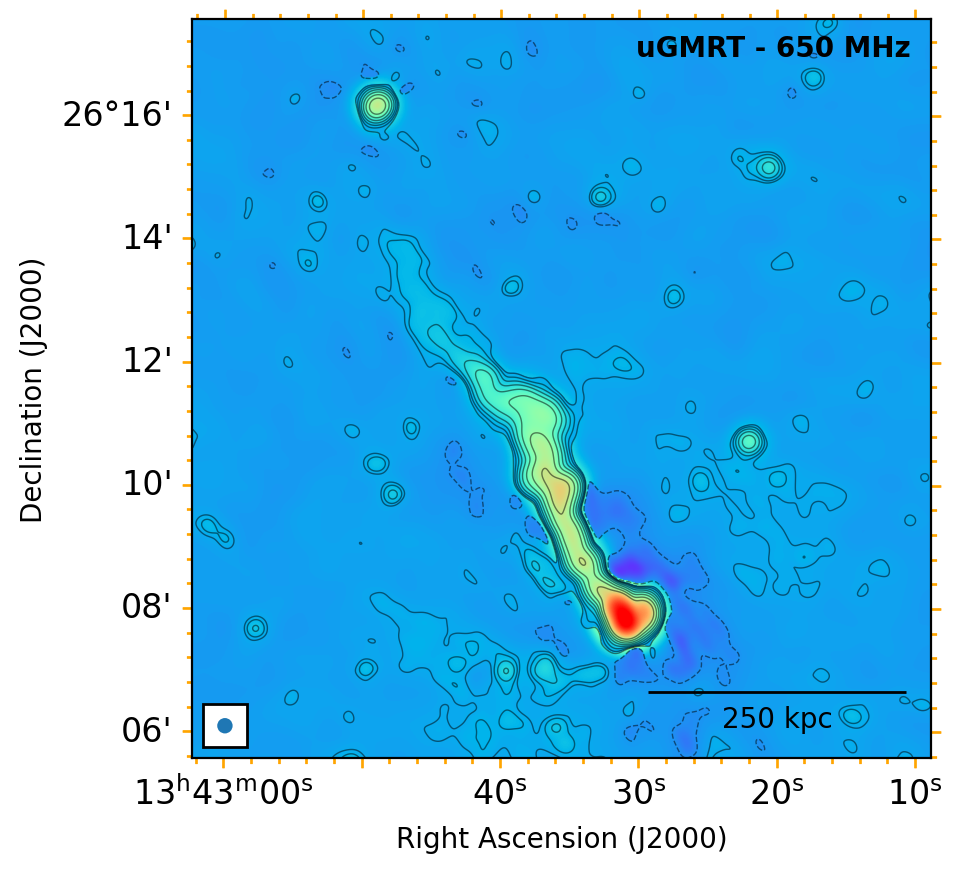}
    \end{subfigure}

    \caption{High- and low-resolution radio images for the central region of Abell 1775 with contours of the head-tail radio galaxy and the revived fossil plasma overlaid. The three images in the first row are high-resolution images with an angular resolution of $6" \times 6"$. The three images in the second row are low-resolution images with an angular resolution of $15" \times 15"$. The noise levels for each image are shown in Tab. \ref{tab:Noise}. All images have the contour levels shown in black drawn at [-1,1,2,4,8,16...]$\times 3\sigma_{rms}$. The colour scale has a logarithmic stretch. Beams are shown in the bottom left corners, and the scale bar is at the bottom right.}
    \label{HighandLowResolutionFinalImages}
\end{figure*}

Figure \ref{HighandLowResolutionFinalImages} shows the multi-frequency view of the radio emission from the central region of A1775 from the images at $6\arcsec$ and $15\arcsec$ resolution that we used to carry out the scientific analysis. The previous study by \cite{Botteon_2021} detected the outer tail of the head-tail radio galaxy only at 144 and 235 MHz. This narrow frequency span between the two observations did not allow for a detailed spectral analysis. However, the use of new data from the uGMRT, with higher resolution and improved sensitivity, enables us to perform such an analysis. All images seen in Fig. \ref{HighandLowResolutionFinalImages} reveal the outer tail, including the high-frequency uGMRT image at 650 MHz. This enables us to cover the radio emission for the entire length of the tail in all images. In low-resolution radio images, the sensitivity of extended emission is enhanced; hence, the tail appears more extended than in the high-resolution images.

The structure of the revived fossil plasma, the filaments F1 and F2, was observed solely with LOFAR from the \cite{Botteon_2021} study. However, as is evident in Fig \ref{HighandLowResolutionFinalImages}, these structures are well constrained in both LOFAR and uGMRT band 3 images, whereas in uGMRT band 4 they are not well detected. 

Since \cite{Botteon_2021} did not detect the outer region of the tail with the VLA at 1.4 GHz, this observation has not been used in this work. We note that \cite{Botteon_2021} also report the presence of central diffuse emission bounded by the cold front, detected only with LOFAR at low resolution ($\sim$30"). In this work, we focus on the head-tail and revived fossil plasma sources, for which it is possible to perform a low-frequency analysis at relatively high resolution (the noise of the uGMRT images significantly increases at resolutions lower than the ones presented here, making the study of the central diffuse emission not suitable).

\subsection{Spectral analysis}

\begin{figure*}[h!]
    \centering
    \captionsetup{labelfont=bf} 
    \begin{subfigure}{0.4\textwidth}
        \centering
        \includegraphics[width=1.2\textwidth]{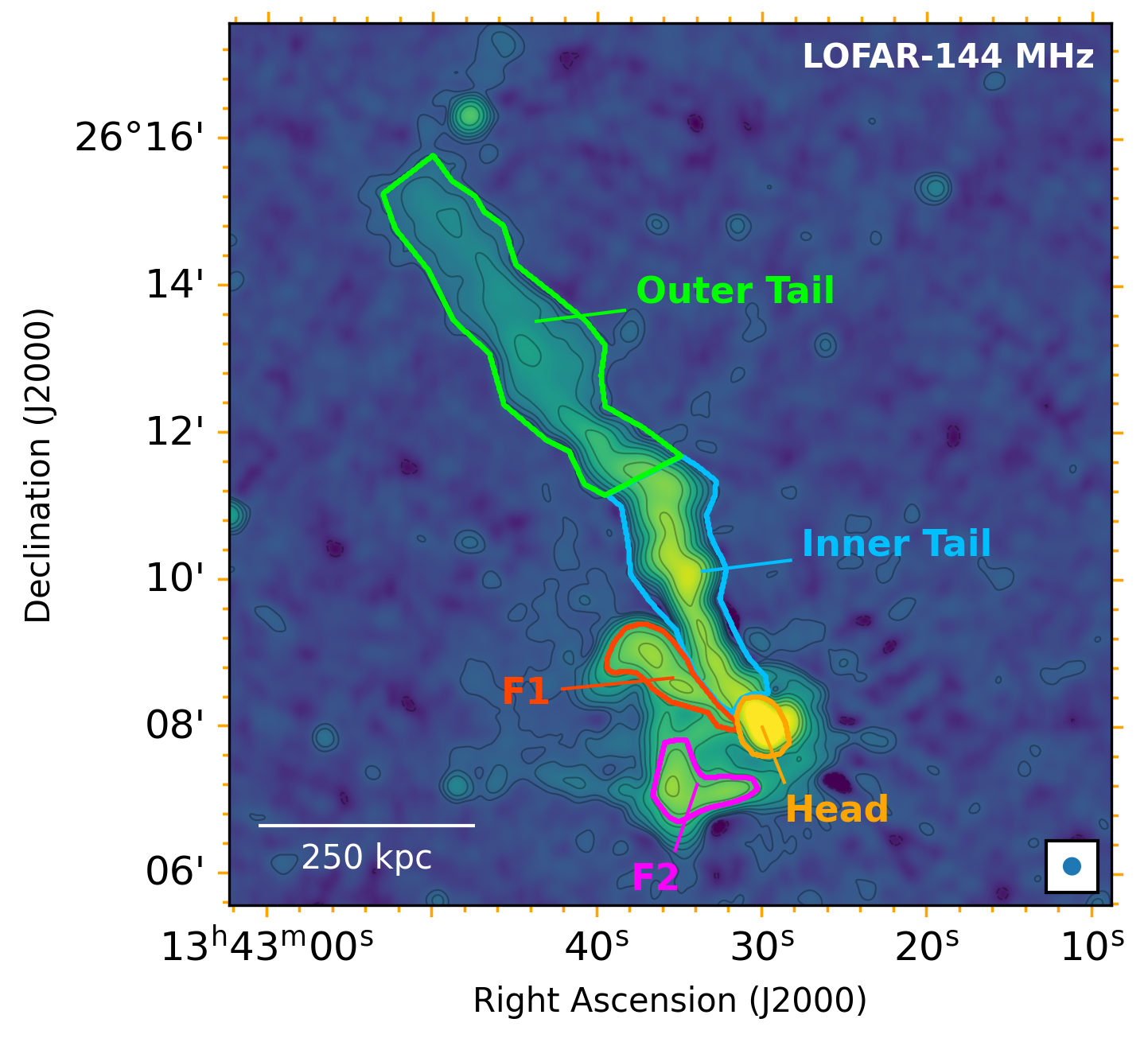}
    \end{subfigure}%
    \hfill
    \begin{subfigure}{0.4\textwidth}
        \centering
        \leftskip -2.0cm
        \includegraphics[width=1.3\textwidth]{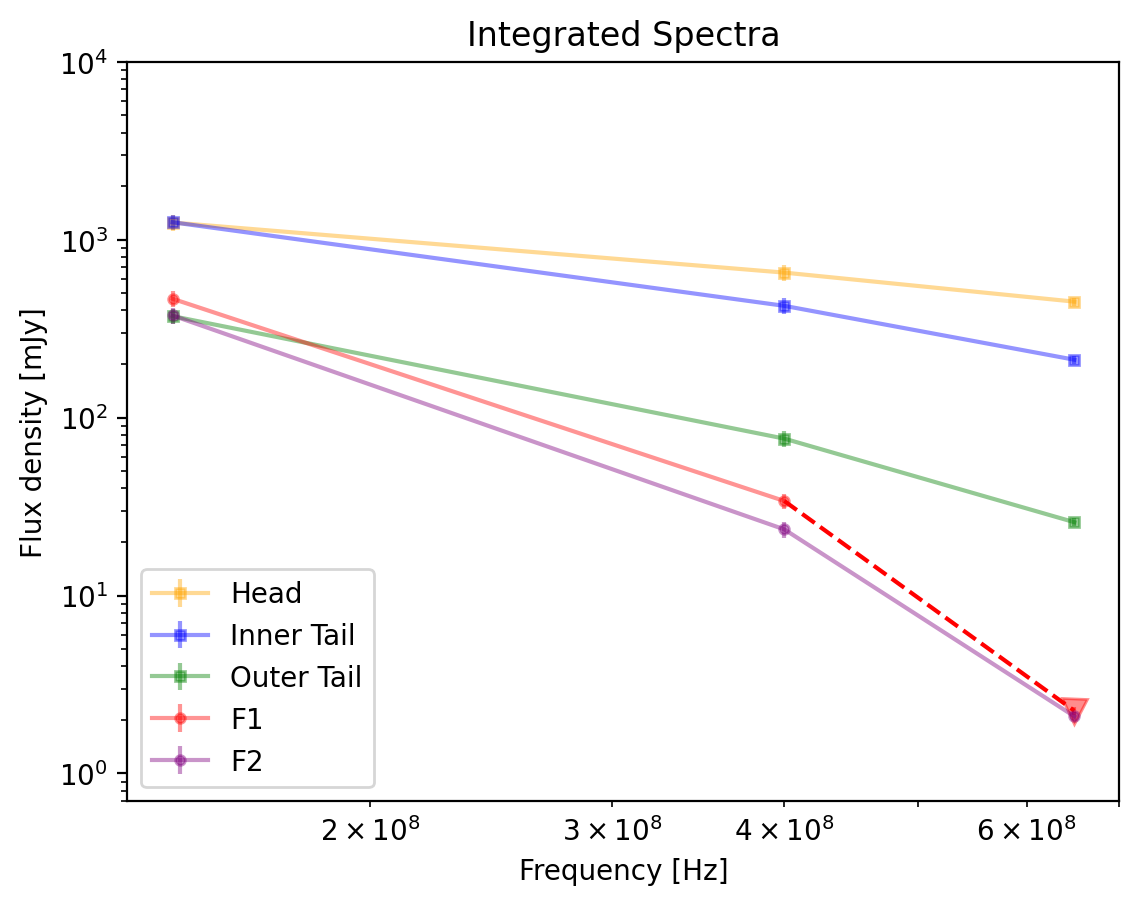}
    \end{subfigure}%
    
\caption{Low-resolution LOFAR radio image at 144 MHz displaying labelled components together with their overlaid regions where the flux density was computed. On the right, we have the integrated spectra of the 'head' and the 'inner' and 'outer' tail components of the head-tail radio galaxy, as well as for the revived fossil plasma, the two filaments F1 and F2.}
    \label{Integrated Flux}
\end{figure*}

We measured the flux density of the head-tail radio galaxy components -- the head and the inner and outer parts of the tail -- by selecting regions from the low-resolution images, as is seen in the left image of Figure \ref{Integrated Flux}. For filament F1 in the uGMRT band 4, due to some artefacts and its structure not being fully detected, we adopted an upper limit on the emission in this band. Filament regions are depicted in the same figure. The integrated flux densities were determined for each frequency and the integrated spectra are displayed on the right of Figure \ref{Integrated Flux}. In Table \ref{IntegratedTable}, we report the flux density values for each frequency and component, along with the calculated spectral indices at low (144-400 MHz) and high (400-650 MHz) frequencies.

The spectral index of the head component is consistent at both high and low frequencies with a single power law.
In the inner tail, there is a more profound steepening between 400 and 650 MHz. The spectral index of the outer tail between 144 MHz and 400 MHz is better constrained compared to \cite{Botteon_2021} because of the wider frequency span in the observations. The outer tail spectral index at high frequency indicates that the spectrum of this component has a strong spectral steepening.
Filaments F1 and F2 were not well detected with uGMRT at 650 MHz. However, the spectral index between 144 and 400 MHz for both filaments is $\alpha_{144}^{400} > 2$, indicating an ultra-steep spectrum.

\begin{table}[h!]
  \captionsetup{labelfont=bf} 
  \centering
  \setlength{\tabcolsep}{3pt} 
  \caption{Flux densities and spectral indices for the A1775 radio sources.}
  \label{tab:TableIntegrated}
  \begin{tabular}{
    |c|c|c|c|c|
  }
    \toprule
   \multicolumn{1}{|c|}{\text{Source}} & \multicolumn{1}{c|}{\text{Frequency}} & \multicolumn{1}{c|}{\text{Flux density}} & \multicolumn{1}{c|}{\text{Spectral Index}} \\ 
   & \multicolumn{1}{c|}{\text{[MHz]}} & \multicolumn{1}{c|}{\multirow{1}{*}{\text{[mJy]}}} & \multicolumn{1}{c|} \phantomsection \\
    \midrule
                 Head & 144 & 1249 $\pm$ 125 & $\alpha_{144}^{400}=0.63 \pm 0.14$ \\
                       & 400 & 653 $\pm$ 65 & $\alpha_{400}^{650}=0.77 \pm 0.23$ \\
                       & 650 & 448 $\pm$ 22 &  \\
            Inner Tail & 144 & 1253 $\pm$ 125 & $\alpha_{144}^{400}=1.06 \pm 0.14$ \\
                       & 400 & 424 $\pm$ 42 & $\alpha_{400}^{650}=1.44 \pm 0.23$ \\
                       & 650 & 211 $\pm$ 10 &  \\
            Outer Tail & 144 & 372 $\pm$ 37 & $\alpha_{144}^{400}=1.55 \pm 0.14$ \\
                       & 400 & 76 $\pm$ 7.6 & $\alpha_{400}^{650}=2.23 \pm 0.23$ \\
                       & 650 & 26 $\pm$ 1.3 &  \\
                    F1 & 144 & 466 $\pm$ 46 & $\alpha_{144}^{400}=2.56 \pm 0.14$ \\
                       & 400 & 34 $\pm$ 3.4 & $\alpha_{400}^{650}>5.59$ \\
                       & 650 & $<$ 2.2 & \\
                    F2 & 144 & 375 $\pm$ 37 & $\alpha_{144}^{400}=2.70 \pm 0.14$ \\
                       & 400 & 23 $\pm$ 2.3 & $\alpha_{400}^{650}=4.98 \pm 0.23$ \\
                       & 650 & 2.1 $\pm$ 0.11 &  \\
    \bottomrule
  \end{tabular}
  \tablefoot{These measurements were performed at the three frequencies 144, 400, and 650 MHz, across the 'head', 'inner' tail, and 'outer' tail of the head-tail radio galaxy, as well as filaments F1 and F2. Spectral index values at high and low frequencies were derived from the integrated spectrum.}
  \label{IntegratedTable}
\end{table}

\begin{figure*}[h!]
    \centering
    \captionsetup{labelfont=bf}
    
    \begin{subfigure}{0.4\textwidth}
        \centering
        \includegraphics[width=1.0\textwidth]{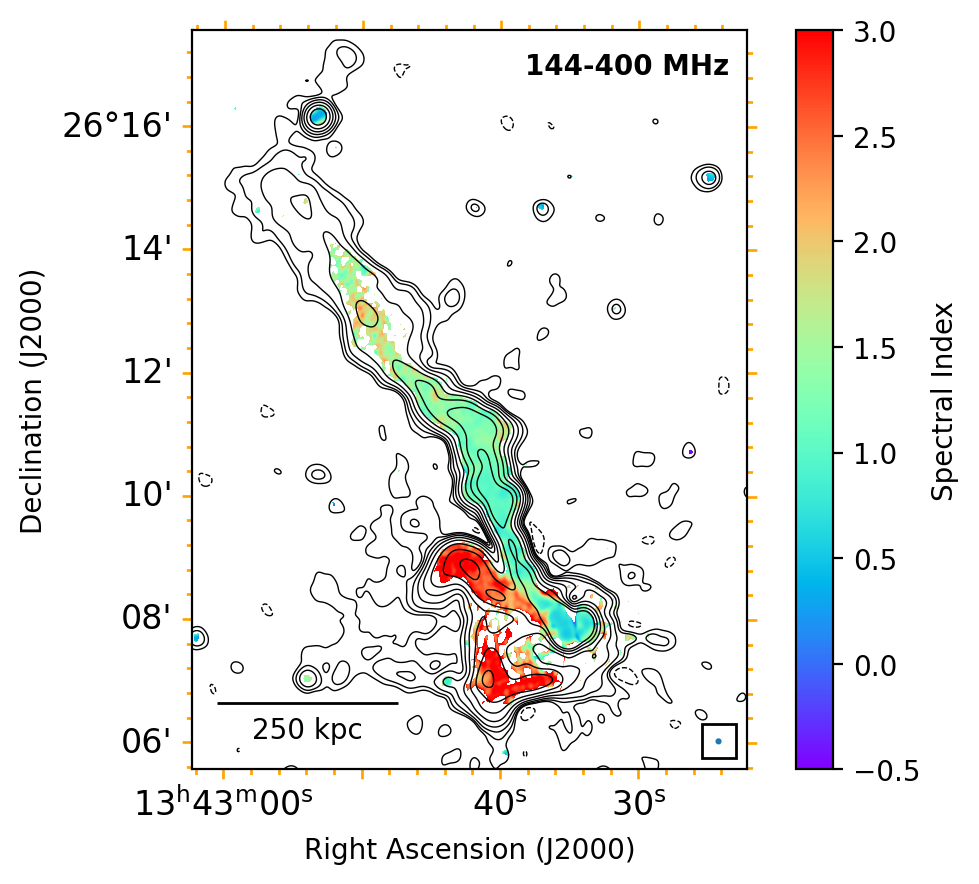}
    \end{subfigure}
    \begin{subfigure}{0.4\textwidth}
        \centering
        \includegraphics[width=1.0\textwidth]{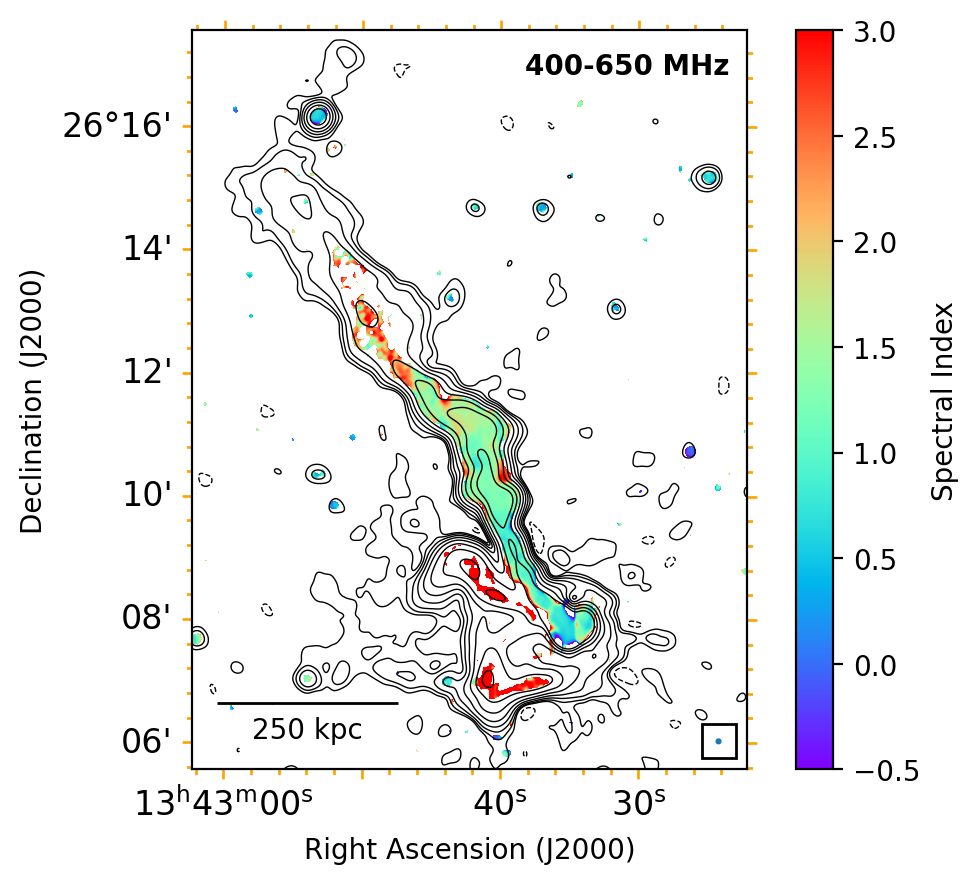}
    \end{subfigure}

    \vspace{0.5cm} 
    \begin{subfigure}{0.4\textwidth}
        \centering
        \includegraphics[width=1.0\textwidth]{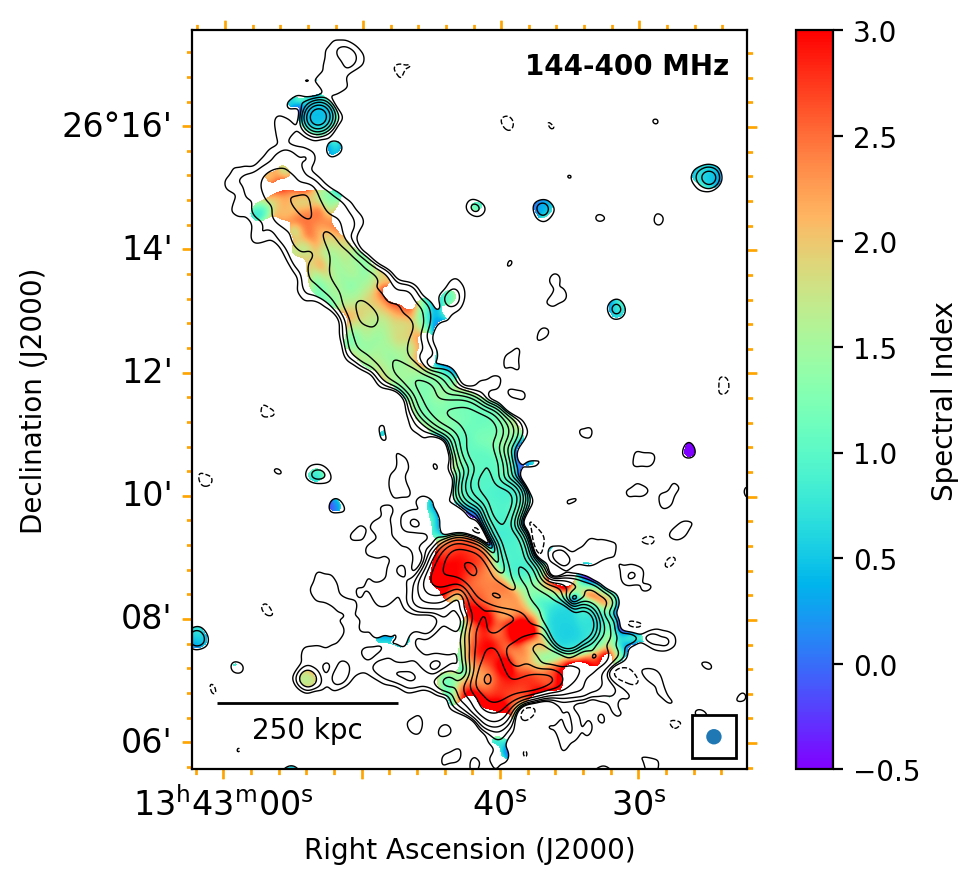}
    \end{subfigure}
    \begin{subfigure}{0.4\textwidth}
        \centering
        \includegraphics[width=1.0\textwidth]{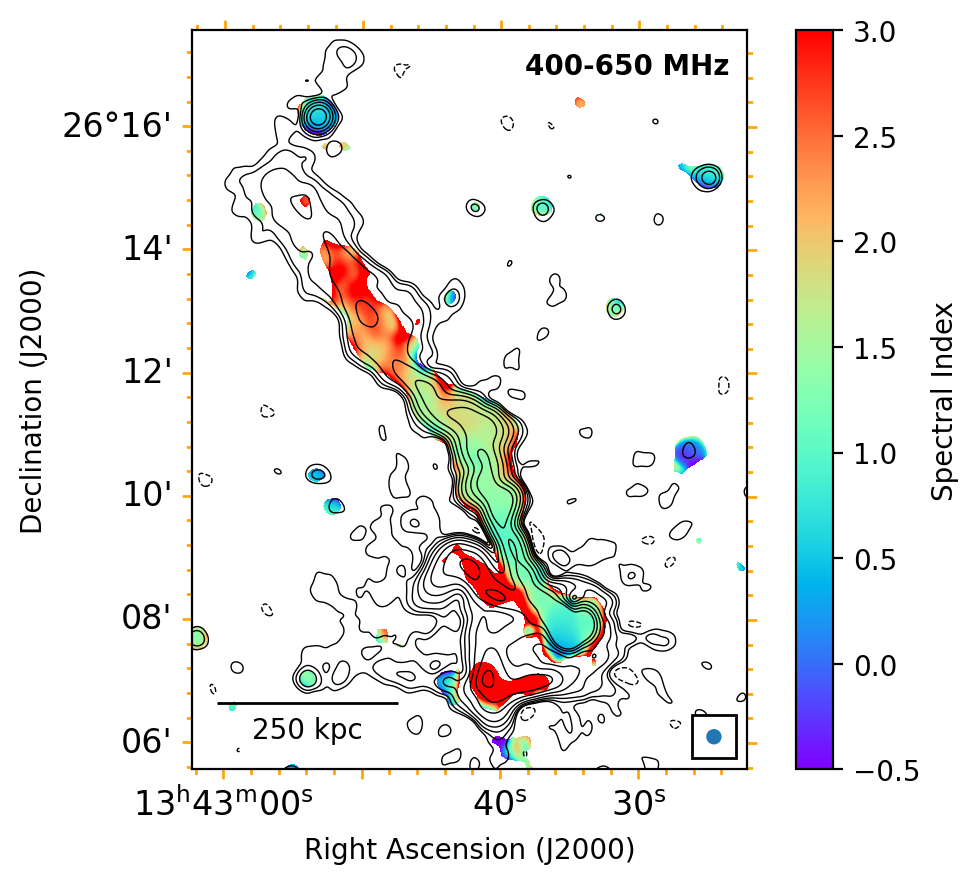}
    \end{subfigure}
    
    \caption{Spectral index maps of the central region of Abell 1775. Upper panel: Spectral index maps of $6'' \times 6''$ resolution radio images. Left: Low-frequency spectral index map between LOFAR and uGMRT band 3. Right: High-frequency map between uGMRT band 3 and 4. Bottom panel: Spectral index maps at $15'' \times 15''$ resolution for both low (on the left) and high frequency (on the right). All images are overlaid with LOFAR contours of 15" resolution. The beam size and the scale bar are indicated in all images.}
    \label{SpectralIndexMaps}
\end{figure*}

Using our multi-frequency data for the central region of Abell 1775, we obtained spectral index maps and corresponding error maps for both high- and low-resolution images. In Figure \ref{SpectralIndexMaps}, we present the low- and high-resolution spectral index maps in the 144-400 MHz and 400-650 MHz frequency range. The high-resolution maps were created to better isolate the contribution of each component's emission, whereas the low-resolution maps were used to capture emission with low surface brightness. Across all maps, we discarded all pixels with a surface brightness significance below the $3\sigma$ level.

These spectral index maps are useful for studying the ageing of relativistic electrons in synchrotron-emitting sources. In particular, low-frequency spectral mapping allows us to trace the emission with steeper spectra that cannot be detected at higher frequencies. 

We note that all maps in Fig. \ref{SpectralIndexMaps} show typical values at the head of $\alpha = 0.5 - 0.8$, and exhibit a spectral steepening along the tail. The low-frequency spectral index maps outline the revived fossil plasma, suggesting an ultra-steep spectrum for these two filaments with $\alpha>2.5$. The corresponding spectral index uncertainty maps are shown in Appendix \ref{A}.

To further probe the spectral shape of the emission, the spectral curvature (SC) maps were obtained using the definition based on three frequencies, derived as follows:

\begin{equation}
     SC={\alpha_{650}^{400}-\alpha_{400}^{144}}  \ .
     \label{Curvature}
\end{equation}

The corresponding uncertainty in SC is defined as:

\begin{equation}
    \Delta SC=\sqrt{(\Delta\alpha_{400}^{144})^2+(\Delta\alpha_{650}^{400})^2} \ .
    \label{CurvatureError}
\end{equation}

\begin{figure*}[h!]
    \centering
    \captionsetup{labelfont=bf}
    
    \begin{minipage}{0.70\textwidth} 
        \centering
        \begin{subfigure}{0.48\textwidth}
            \centering
            \includegraphics[width=\linewidth]{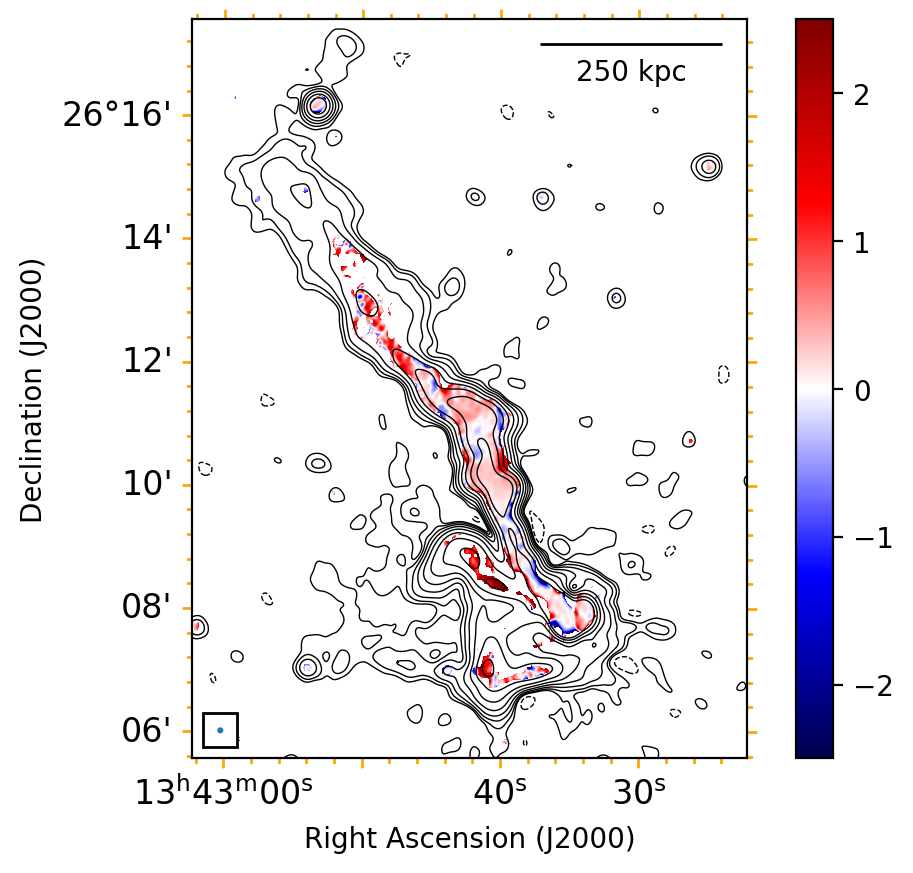}
        \end{subfigure}%
        \begin{subfigure}{0.48\textwidth}
            \centering
            \includegraphics[width=\linewidth]{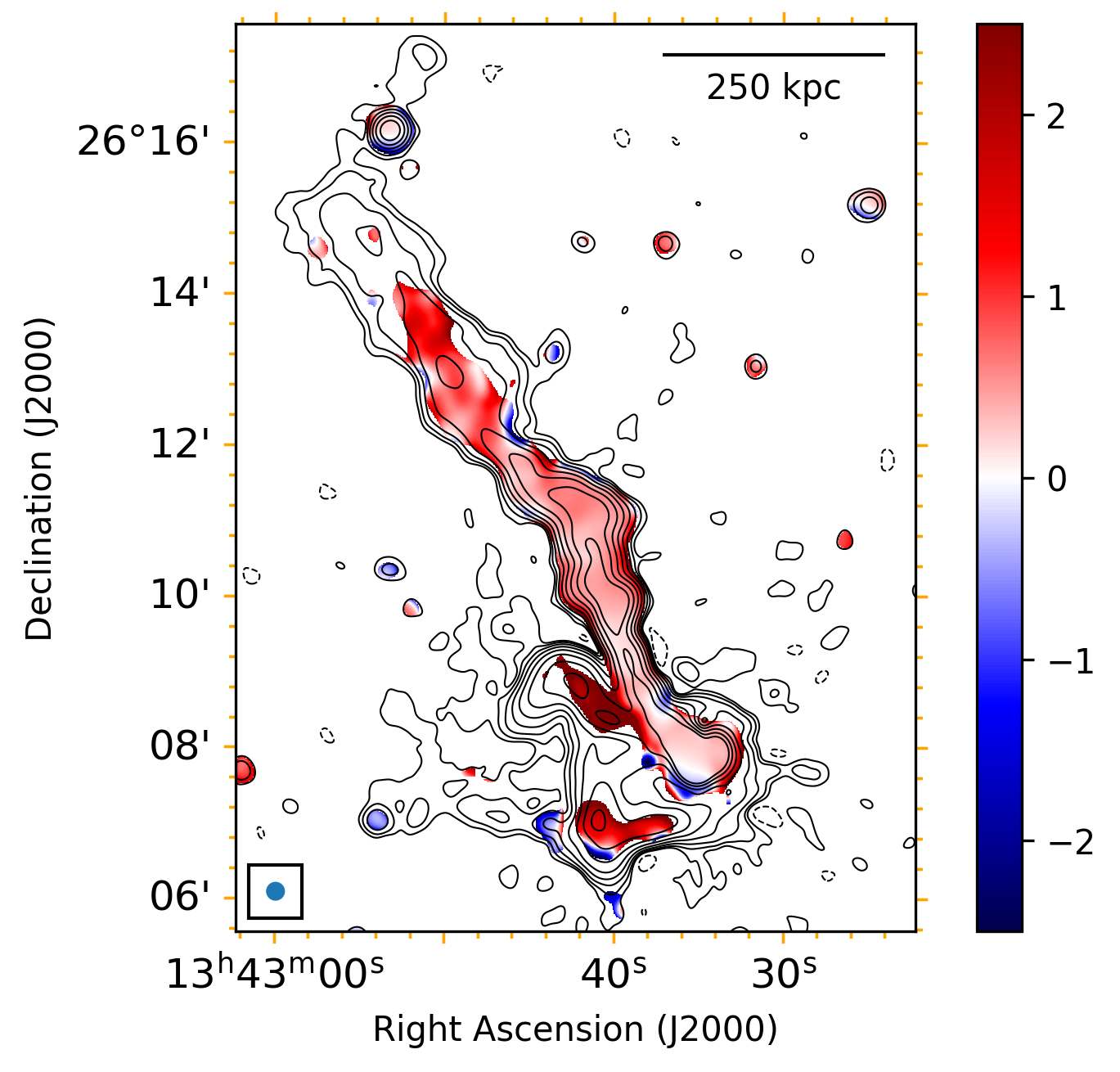}
        \end{subfigure}
    \end{minipage}%
    \hfill
    \begin{minipage}{0.3\textwidth} 
        \caption{Spectral curvature maps of the central region of Abell 1775. Left: Spectral curvature map of high-resolution radio images ($6"\times 6"$). Right: Spectral curvature map of low-resolution radio images ($15"\times 15"$). Both maps are three-frequency SC maps between 144 MHz and 650 MHz. Contour levels are drawn at [-1,1,2,4,8,16...]$\times 3\sigma_{rms}$ of the LOFAR image at 15" resolution. The diverging colour bar shows the SC from -2 to 2, where the white colour is when SC=0.}
        \label{CurvatureMap}
    \end{minipage}

\end{figure*}

The SC maps in high and low resolution are displayed in Figure \ref{CurvatureMap}, where the positive and negative values correspond to convex and concave spectra, respectively. The head-tail radio galaxy exhibits a spectral index steepening along the tail with some local fluctuations, which is enhanced in the high-resolution map.
The corresponding SC uncertainty maps are shown in Appendix \ref{A}.

\subsection{Head-tail radio galaxy \label{3.2}}

\begin{figure}[h!]
    \centering
    \captionsetup{labelfont=bf}
    \begin{subfigure}{0.30\textwidth}
        \centering
        \includegraphics[width=1.1\textwidth]{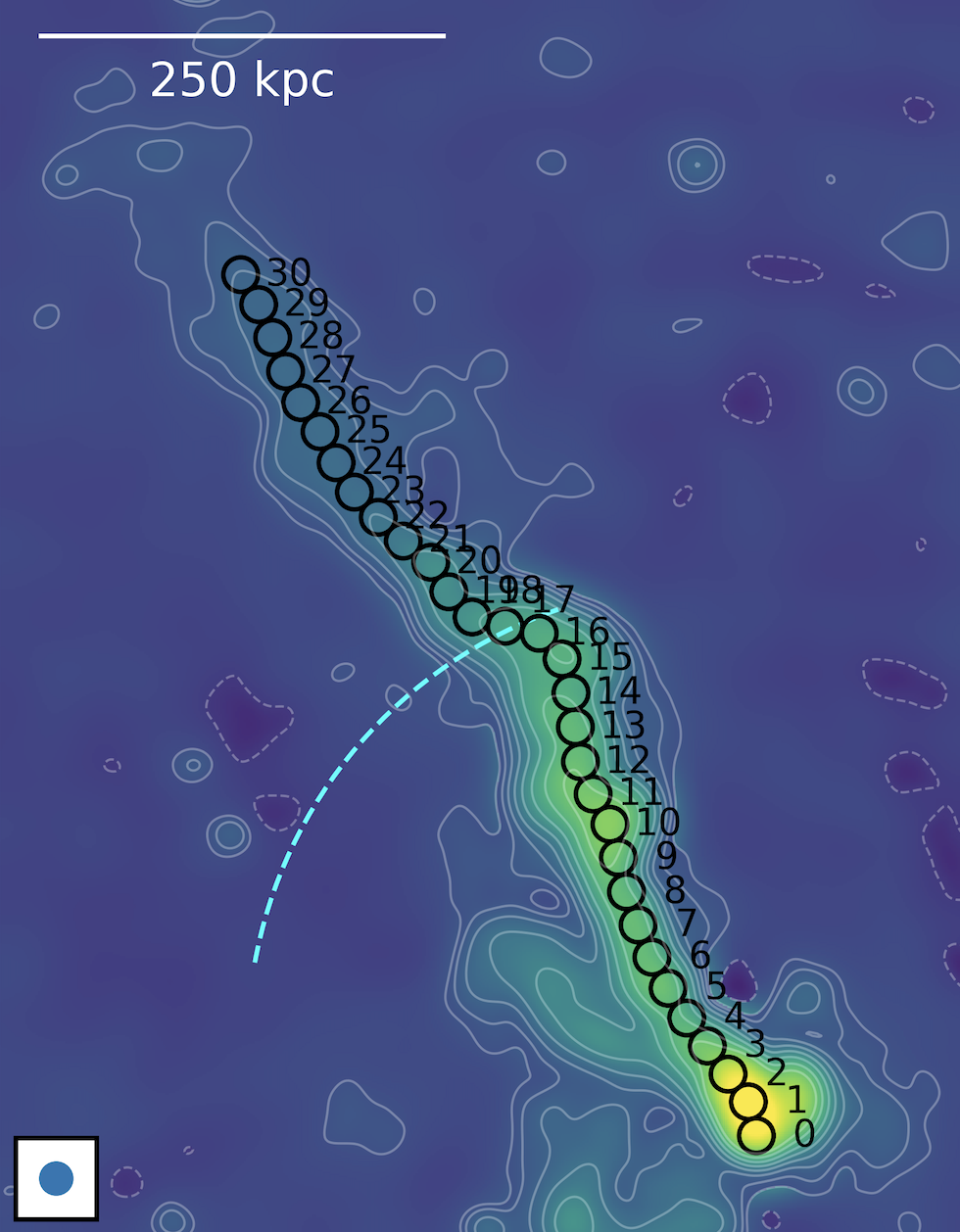}
    \end{subfigure}
    \hfill    
    \caption{Circular beam-sized regions with 15\arcsec diameter drawn based on low-resolution images, using a $3 \sigma$ emission threshold to identify the outermost region. They start from 0, marking the head of the tail, and extend up to 30, covering the entire length of the tail. These regions are overlaid in a uGMRT band 3 low-resolution image, with the arc-shaped cold front shown with a dashed blue line. The beam is shown in the bottom left corner, together with the scale bar in the top left corner.}
    \label{Head-tail-regions}
\end{figure}

\begin{figure}[h!]
    \centering
    \captionsetup{labelfont=bf}
    \hspace*{-0.2cm}
        \begin{subfigure}{0.5\textwidth}
        \centering
        \includegraphics[width=1.0\textwidth]{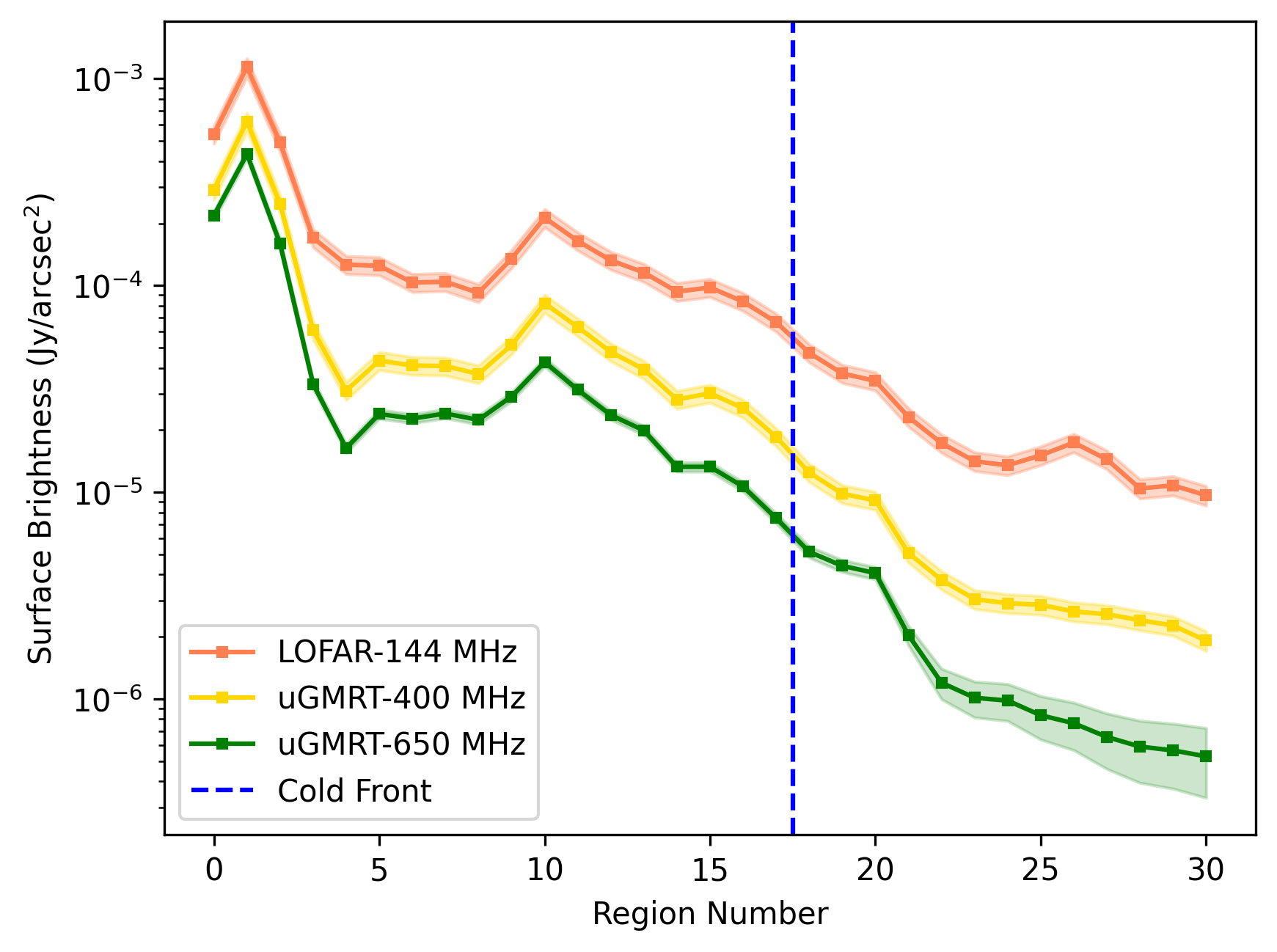}
    \end{subfigure}
    \hfill    
    \begin{subfigure}{0.5\textwidth}
        \centering
        \includegraphics[width=1.0\textwidth]{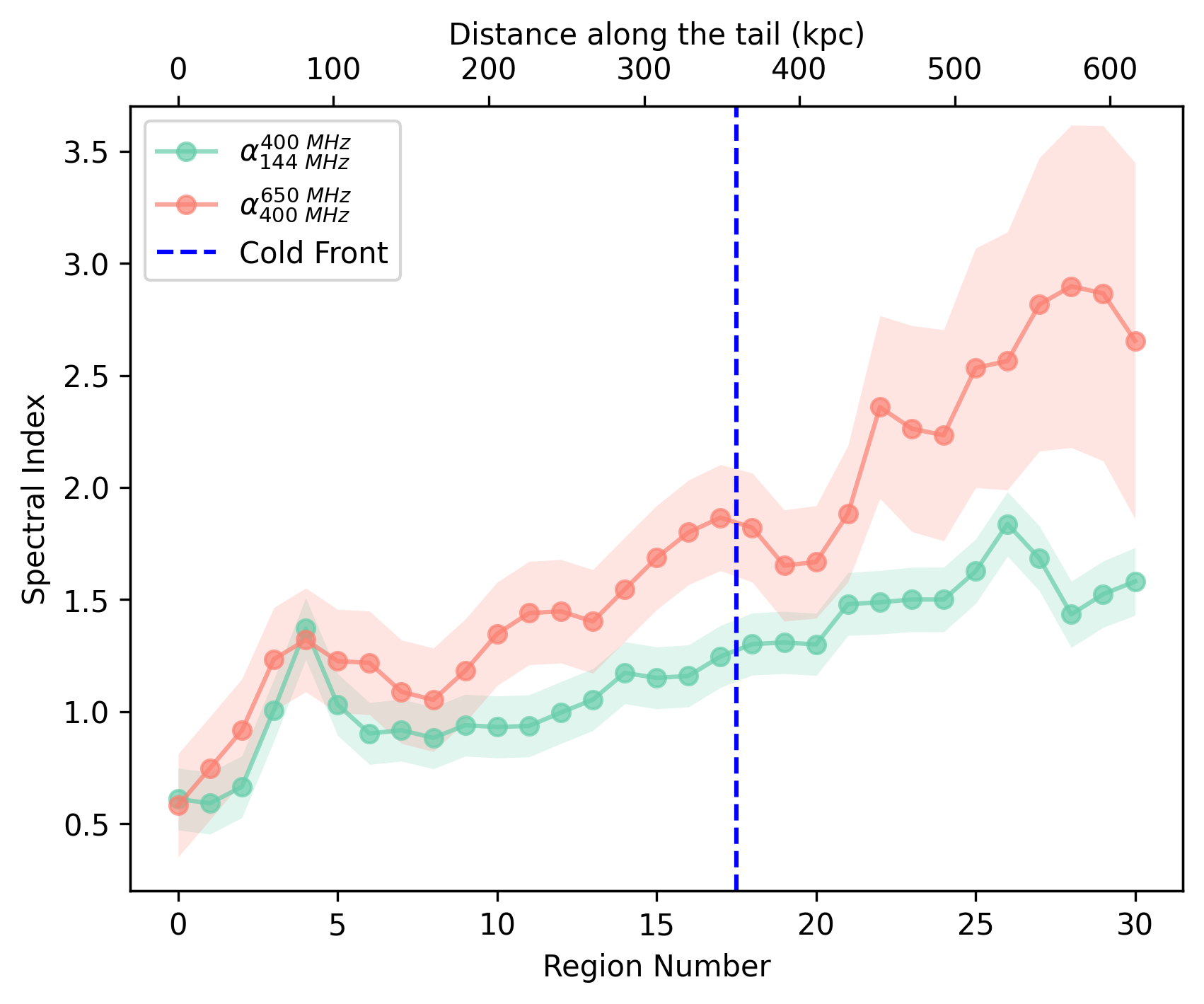}
    \end{subfigure}
    \hfill    
    \caption{Surface brightness and spectral index profiles for the head-tail radio galaxy. Upper panel: Surface brightness trends along the head-tail measured from low-resolution images of the LOFAR at 144 MHz, followed by the uGMRT at 400 MHz, and the uGMRT at 650 MHz. Bottom panel: Spectral index profiles of low-resolution images of low and high frequency, over region numbers and projected distance along the tail in kiloparsecs. Distances in kiloparsecs were calculated, starting from region 0. The flux density for both plots was extracted from circular regions, as is depicted in Fig. \ref{Head-tail-regions}. The vertical dashed blue line in both plots denotes the location of the arc-shaped cold front, where the tail 'breaks', as can be seen in Fig. \ref{fig:Head-tail-Botteon}.}
    \label{SpectralIndexPlots}
\end{figure}

In order to study the surface brightness profile observed along the head-tail radio galaxy structure in more detail, we produced the plot reported in the upper panel of Fig. \ref{SpectralIndexPlots}. In this plot, we compare the surface brightness profiles of LOFAR at 144 MHz and the uGMRT at 400 MHz and 650 MHz at low resolution. We drew circular regions on the images starting from 0, corresponding to the southern tip of the head, where the flux density was evaluated. These regions, each the size of a beam of the low-resolution image, are shown in Figure \ref{Head-tail-regions}. We used 30 circular regions that follow the head-tail ridge line where the emission exceeds  $\> 3\sigma$ in all observations.

The surface brightness profiles of all low-resolution images reveal that within the innermost regions (0 to 2), corresponding to the head of the tail, the radio surface brightness is the highest. Notably, at region 1, the surface brightness reaches its peak, with values of $1.14 \pm 0.11 \times 10^{-3} \ \text{Jy/}{\text{arcsec}^{2}}$ in  LOFAR, of $6.22 \pm 0.62 \times 10^{-4} \ \text{Jy/}{\text{arcsec}^{2}}$ in uGMRT band 3, and of $4.33 \pm 0.22 \times 10^{-4} \ \text{Jy/}{\text{arcsec}^{2}}$ in uGMRT band 4. Then the emission sharply decreases until it reaches region 4 (corresponding to $\sim 82 \ \text{kpc}$ in the projected distance along the tail) and remains roughly constant until region 8 (corresponding to $\sim 164 \ \text{kpc}$). A more pronounced brightness peak is observed between regions 8 and 14, corresponding to a knot visible in Fig \ref{Head-tail-regions}. This is followed by a gradual decrease from region 14 onwards, including the regions beyond the cold front (shown with the dashed blue line, as is seen also in Fig.\ref{fig:Head-tail-Botteon}). However, the LOFAR profile shows an increase in surface brightness between regions 24 and 28, unlike the other profiles.
For all three frequencies, the surface brightness has its minimum in the last region at the end of the tail detected above 3$\sigma$.

We computed the low-frequency spectral index profiles between LOFAR at 144 MHz and uGMRT at 400 MHz as well as the high-frequency one between uGMRT at 400 and 650 MHz for both high- and low-resolution images.
In the bottom panel of Fig. \ref{SpectralIndexPlots}, we show the spectral index trends along the tail for low-resolution images using the same circular regions used to compute the surface brightness profiles.

In the low-frequency spectral index profile of the low-resolution images, an unexpected steepening in the spectral index is observed from regions 4 to 8, ranging from $\alpha = 1.37 \pm 0.14$ to $\alpha = 0.88 \pm 0.14$. Following region 8, the profile gradually steepens until region 30, with only a slight flattening in the spectral index observed between regions 24 and 28.

In the high-frequency spectral index profile, the steepening in the spectral index between regions 4 and 8 is less prominent. We suspect that this unexpected steepening in the spectral index in the inner regions of the tail is due to the contamination of the emission from the nearby filament F1. Further details are provided in Section \ref{4}.
Considering the uncertainties represented by the shaded area, the rest of the high-frequency spectral index profile is comparable to the low-frequency one.
Overall, the high-frequency profile exhibits higher values for the spectral index compared to the low-frequency one, reaching a maximum value in the region 28 with $\alpha=2.90\pm0.72$, in line with the increasing SC observed along the tail seen in Fig.\ref{CurvatureMap}.

\subsubsection{Colour-colour diagram \label{3.2.1}}
To identify possible departures from a pure ageing model, we compared radio colour-colour diagrams with three different models such as JP \citep{Jaffe_and_Parola_1973}, KP \citep{Kardashev_1962,Pacholczyk_1970}, and Tribble \citep{Tribble_1993}, which we computed using the \textsc{BRATS}\footnote{\url{https://www.askanastronomer.co.uk/brats/} \label{ft:Bratsfootnote}} software \citep{Harwood_2013, Harwood_2015}. Across all models, we assumed an electron injection spectral index of $\alpha_\text{inj}=0.6$, consistent with the typical spectral index values measured from the first two regions (see the bottom panel of Fig. \ref{SpectralIndexPlots}), yet the models differ in radiative losses. The KP and JP models consider a constant magnetic field, the former with constant electron pitch angles and the latter with their isotropization. For both models, we decided to keep the magnetic field fixed at the minimum ageing value, $B=B_{min}$ (see below).
Tribble, in the JP context, allows variation in the magnetic field.

The magnetic field strength $B_{min}$  corresponds to the field that maximizes the radiative lifetime of electrons: higher or lower values would result in shorter lifetimes. This minimal loss magnetic field was computed as \citep[e.g.][]{Pacholczyk_1970}:

\begin{equation}
    B_{min} [\mu G]=\frac{B_{CMB}}{\sqrt{3}}=\frac{3.25 (1+z)^2}{\sqrt{3}} \ ,
\end{equation}

where $B_{CMB}$ is the equivalent magnetic field strength of the cosmic microwave background radiation.
In our case, the resulting magnetic field strength is $B_{min}\simeq 2.15 \mu G$. Using this value, the evolutionary tracks derived from the ageing models were created and adopted for our head-tail radio galaxy.

\begin{figure}[h!]
    \centering
    \captionsetup{labelfont=bf}
    \begin{subfigure}{0.5\textwidth}
        \centering
        \includegraphics[width=1.0\textwidth]{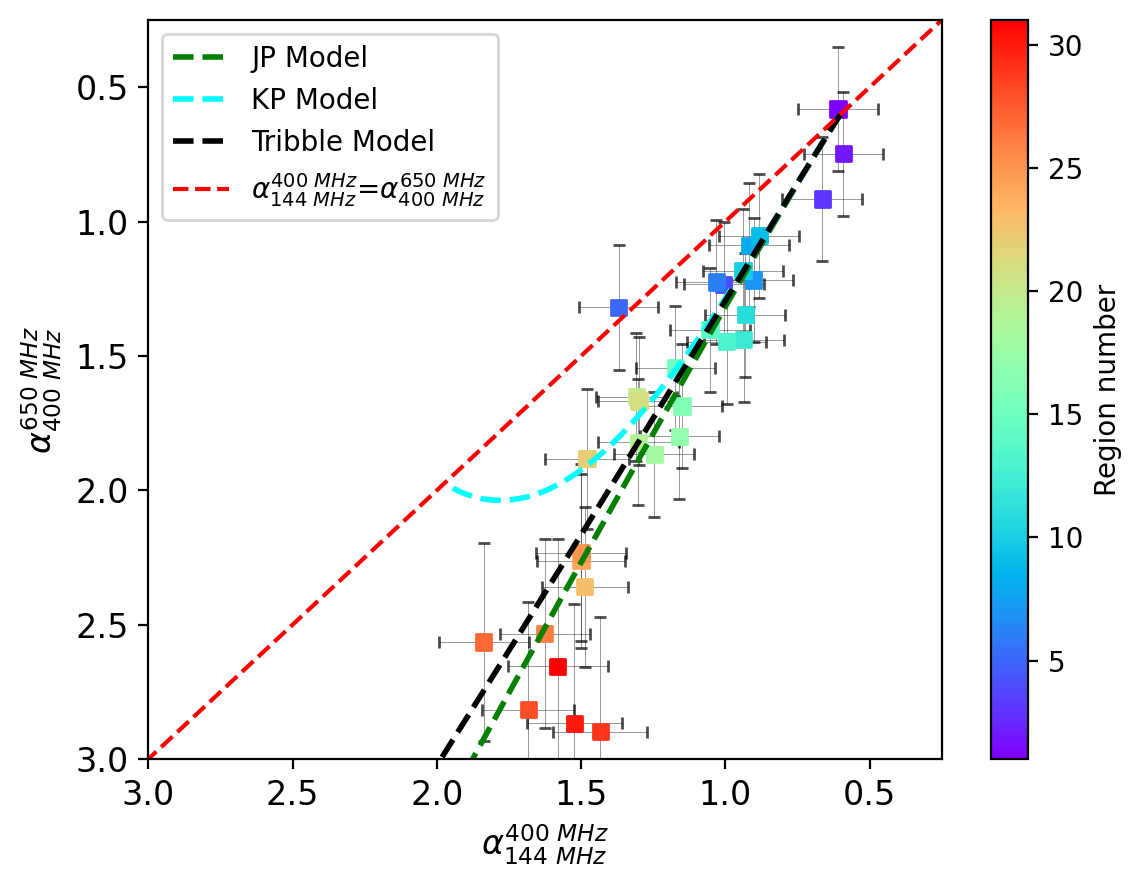}
    \end{subfigure}
    \hfill    
    \caption{Radio colour-colour diagram of low-resolution images with ageing models represented by the dashed green line for the JP model, the dashed aquamarine line for the KP model, and the Tribble model shown in black. The colour bar shows the region number. The injection spectral index adopted for all the models is $\alpha_{\text{inj}}=0.6$.}
    \label{fig:Color-color with model}
\end{figure}

These three-frequency radio colour-colour diagrams were generated to complement spectral analysis along the head-tail, providing similar information to the curvature maps. We computed the colour-colour diagram considering 144-400 and 400-650 spectral indices, using the same 15\arcsec regions as were used in the previous analysis. This diagram is shown in Figure \ref{fig:Color-color with model} together with the JP, KP, and Tribble models. The dashed red line is the unit line, which denotes the power-law behaviour that corresponds to $\alpha_{144}^{400}=\alpha_{400}^{650}$. Any data point that lies away from the unit line indicates curvature. In this diagram, most of the data points are located below the power-law line, indicating a negative curvature (convex spectrum). This colour-colour diagram will be further discussed in Section \ref{4}.
 
\subsubsection{Equipartition magnetic field \label{3.2.2}} 
We estimated the strength of the magnetic field along the tail using the minimum energy conditions.
For minimum energy, we adopted the equation from \cite{Govoni2004} in terms of observed quantities as follows:

\begin{equation}
\begin{split}
u_{min}\left[\frac{\text{erg}}{\text{cm}^{3}}\right] &= \xi(\alpha, \nu_{1},\nu_{2})(1+k)^{4/7}(\nu_{0[\text{MHz}]})^{{4\alpha}/{7}} \\
&\quad \times (1+z)^{{12+4\alpha}/{7}}(I_{0[\frac{\text{mJy}}{\text{arcsec}^{2}}]})^{{4}/{7}}(d_{[\text{kpc}]})^{{-4}/{7}} \ ,
\end{split}
\end{equation}

where $I_0$ is the source brightness at the central frequency, $\nu_0$, and $d$ is the source depth, which in our case was computed as the diameter of the circular regions in kiloparsecs. The constant $\xi(\alpha, \nu_{1},\nu_{2})$ we used was based on the injection spectral index (0.6), and it was $\xi(\alpha, \nu_{1},\nu_{2})=1.55\times 10^{-12}$ for $\nu_1$=10 MHz and $\nu_2$=10 GHz \citep[see][]{Govoni2004}. The ratio of the energy in relativistic protons to that in electrons, $k$, depends on the mechanism by which relativistic electrons are generated, which is so far poorly known. In our case, we assume $k=1$, as is commonly done in the literature \citep{Govoni2004}. As the relativistic plasma ages, the electron energy decreases (due to radiative losses) with respect to that of protons, implying a progressive increase in the value of $k$. Assuming $k=1$, this likely results in an underestimate of the total energy content (and the equipartition magnetic field) as we move along the tail.

The equipartition magnetic field was then obtained from:

\begin{equation}
B_{eq} = \left( \frac{24\pi}{7} u_{min} \right)^{\frac{1}{2}} \ .
\label{B_eq}
\end{equation}

\begin{figure} [h!]
    \centering
    \captionsetup{labelfont=bf}
    \begin{subfigure}{0.5\textwidth}
        \centering
        \includegraphics[width=1.0\textwidth]{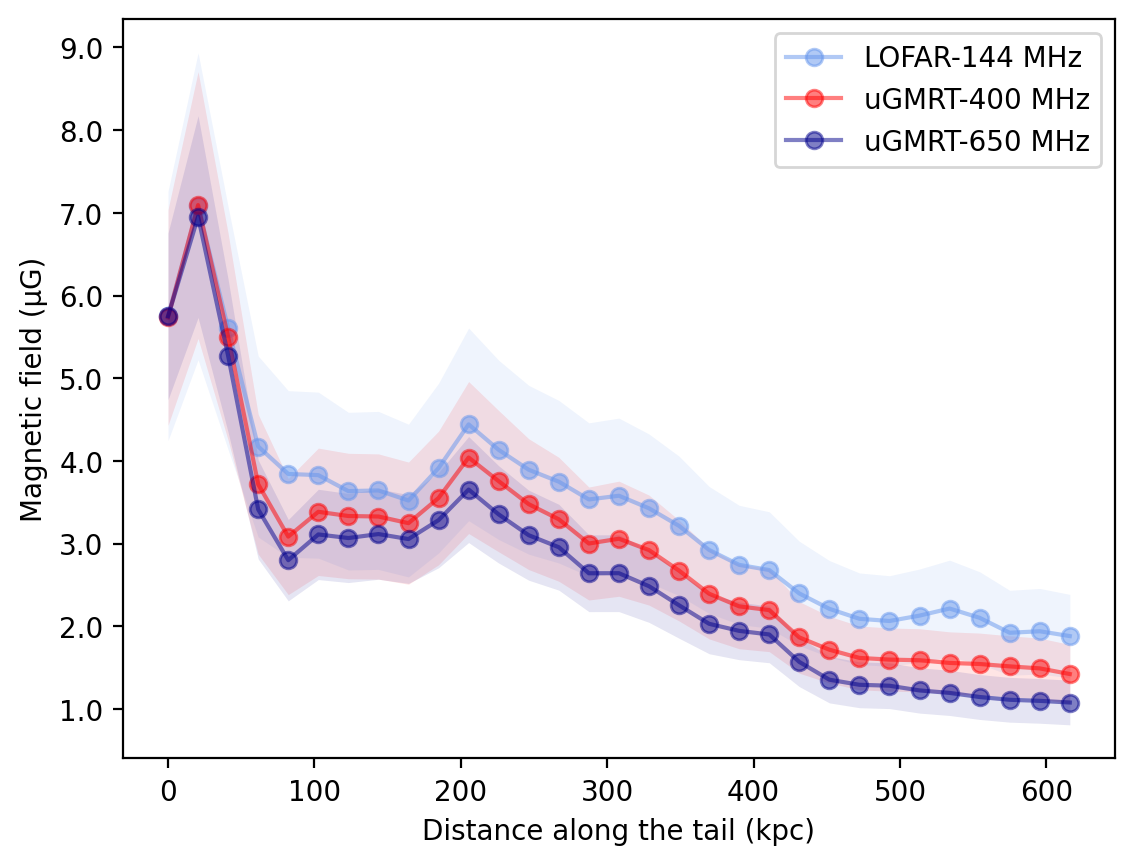}
    \end{subfigure}%
    \caption{Estimated trends of the equipartition magnetic field in microgauss over the projected distance along the tail in kiloparsecs. The profiles were derived from low-resolution images of LOFAR at 144 MHz and uGMRT at 400 MHz and 650 MHz.}
    \label{MagneticField}
\end{figure}

In this standard method, the equipartition parameters are calculated by integrating the synchrotron radio luminosity between the fixed high- and low-frequency cut-off, $\nu_1$ and $\nu_2$. However, this approach is not entirely correct, as the integration limits should vary with the energy of the radiating electrons. Instead, low- and high-energy cut-offs for the particle distribution should be applied \citep{Brunetti1997, Beck2005}. Indicating the electron energy by its Lorentz factor, assuming that $\gamma_{min}<<\gamma_{max}$, the revised equipartition magnetic field strength ($B_{eq}$) is (for $\alpha>0.5$):

\begin{equation}
    B'_{eq}\sim 1.1 \ \gamma_{min}^\frac{1-2\alpha}{3+\alpha} \ B_{eq}^\frac{7}{2(3+\alpha)} \ 
,\end{equation}

where $B_{eq}$ is the value we obtained from the standard formula in Eq.\ref{B_eq}. We assumed $\gamma_{min}$=100 for the lower cut-off. The exact value of the lower cut-off is difficult to estimate, as it depends on the shape of the radio spectrum.
Using the revised equipartition formula, we obtained equipartition magnetic field profiles shown in Fig. \ref{MagneticField}, from low-resolution images of LOFAR at 144 MHz and uGMRT at 400 MHz and 650 MHz.
The equipartition magnetic field exhibits a decreasing trend along the tail in all three profiles depicted in the figure. The errors in the magnetic field profile plot are estimated by minimum energy, considering only the uncertainties in flux density.

\subsection{Revived fossil plasma}
We created plots to search for possible trends along the surface brightness profiles of filaments F1 and F2. The surface brightness measurements were performed only from LOFAR at 144 MHz and uGMRT at 400 MHz observations due to the non-detection of the filaments in uGMRT at 650 MHz, as is evident in Fig. \ref{HighandLowResolutionFinalImages}.

\begin{figure}[h!]
    \centering
    \captionsetup{labelfont=bf}
    \begin{subfigure}{0.28\textwidth}
        \centering
        \includegraphics[width=1.0\textwidth]{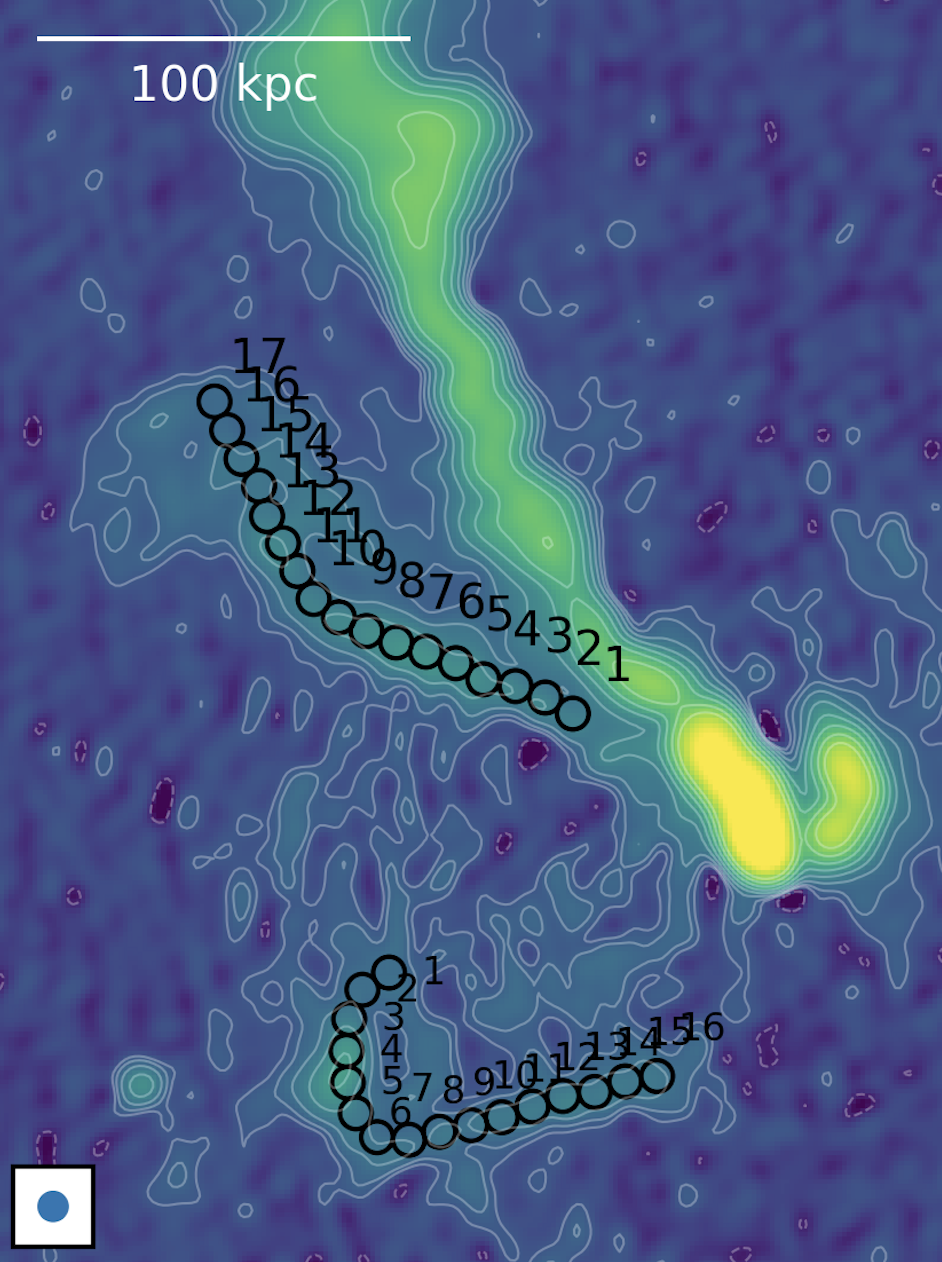}
    \end{subfigure}
    \hfill    
    \caption{Circular beam-sized regions with a diameter of 6\arcsec were drawn based on high-resolution images for both filaments F1 and F2, using a $3 \sigma$ emission threshold. Both filament regions start from 1; F1 has 17 regions, while F2 has 16. These regions are overlaid in the uGMRT band 3 high-resolution image and the beam is shown in the bottom left corner.}
    \label{Filaments-regions}
\end{figure}

\begin{figure*}[h!]
    \centering
    \captionsetup{labelfont=bf}
    
    \begin{subfigure}{0.33\textwidth}
        \centering
        \includegraphics[width=1.0\textwidth]{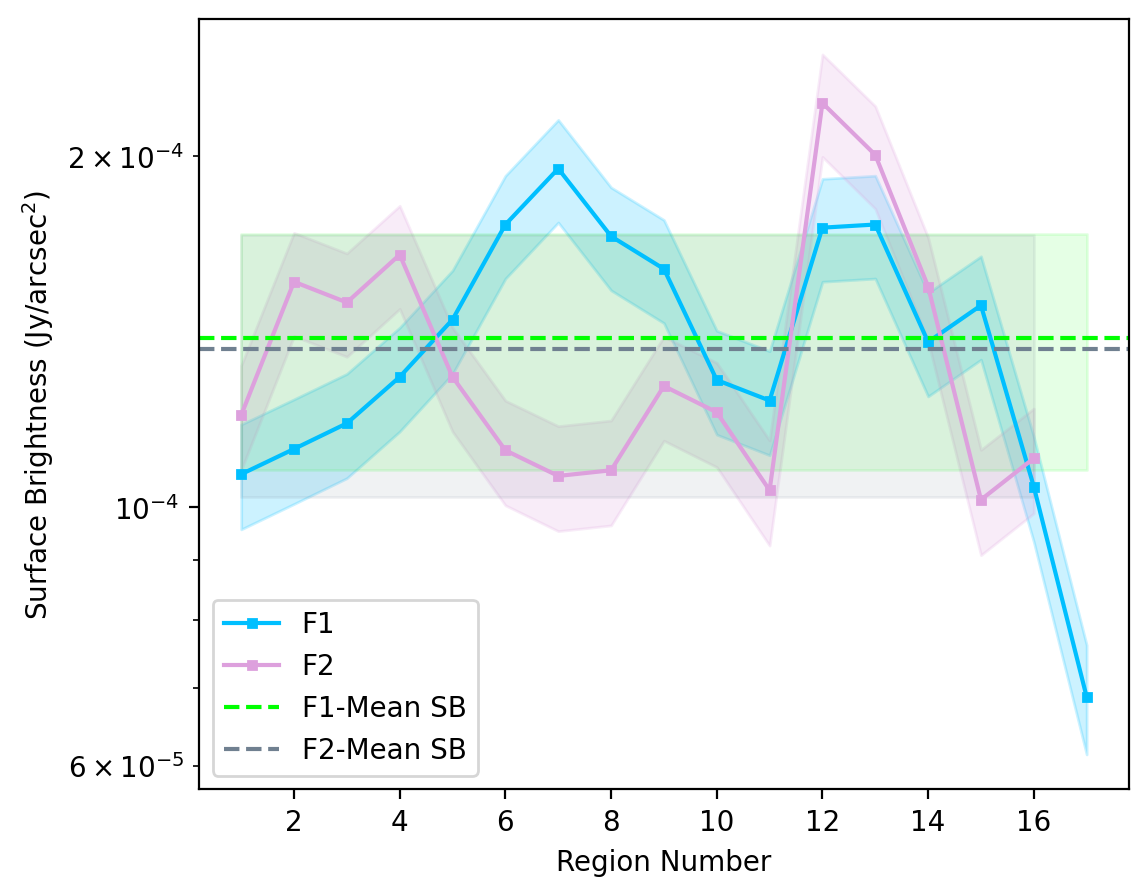}
    \end{subfigure}
    \begin{subfigure}{0.33\textwidth}
        \centering
        \includegraphics[width=1.0\textwidth]{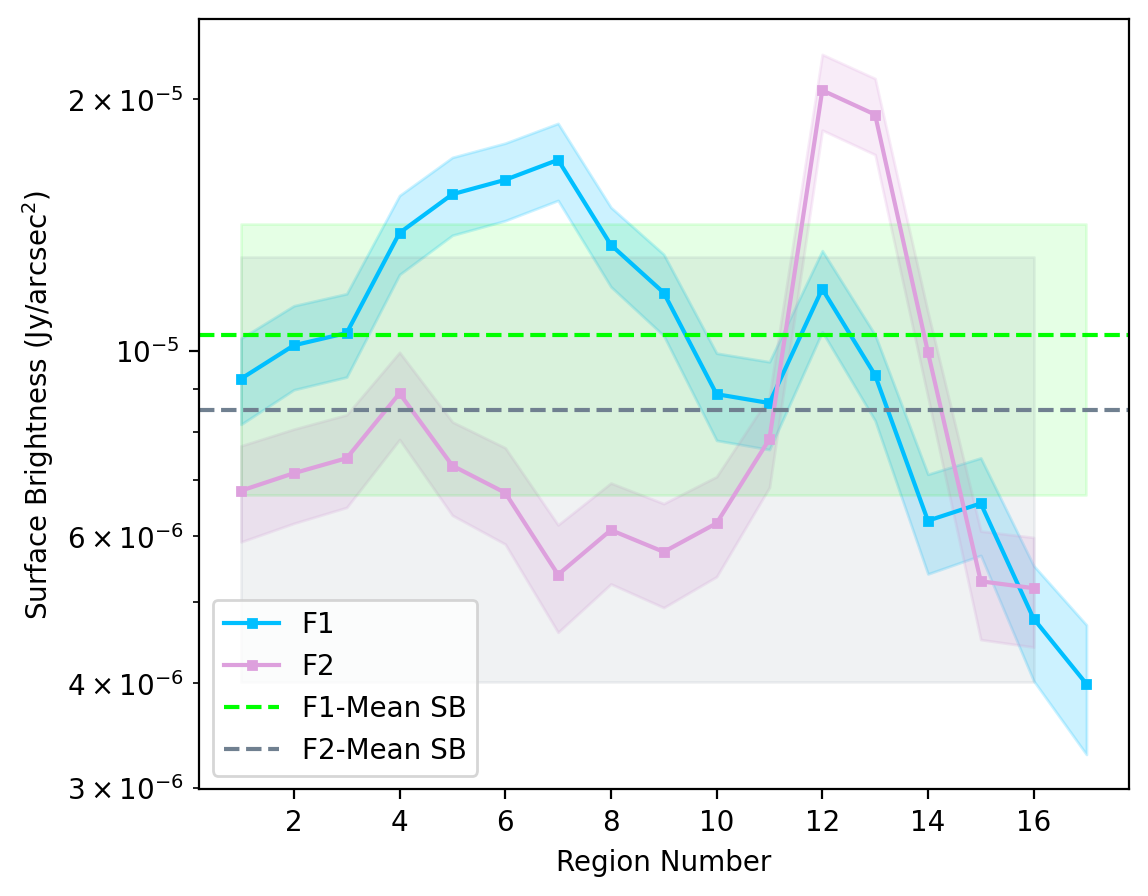}
    \end{subfigure}
    \begin{subfigure}{0.33\textwidth}
        \centering
        \includegraphics[width=1.0\textwidth]{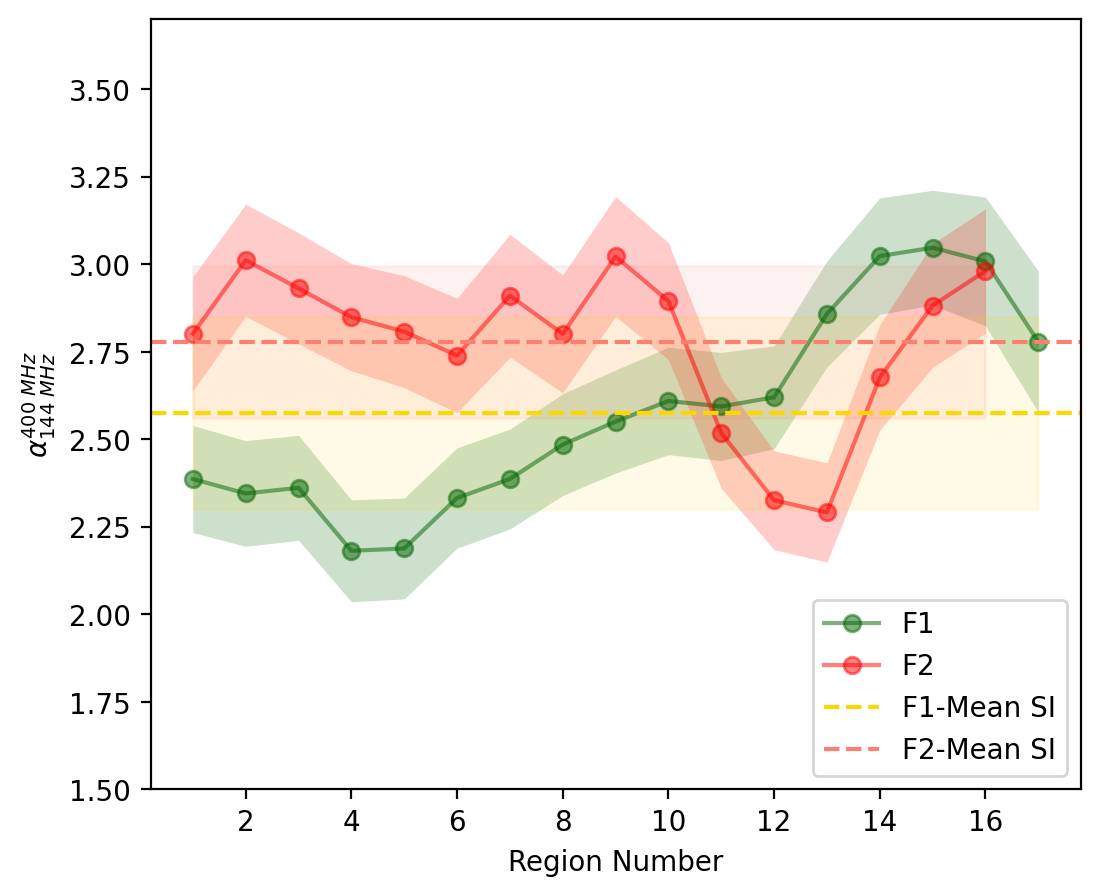}
    \end{subfigure}
    
    \caption{High-resolution surface brightness and spectral index trends for filaments F1 and F2. The left panel shows the surface brightness profiles of the LOFAR image at 144 MHz, followed by the uGMRT image profiles at 400 MHz in the middle panel. In the right panel, we show the spectral index profiles of filaments F1 and F2, computed from high-resolution images between LOFAR at 144 MHz and uGMRT at 400 MHz. The horizontal dashed lines represent the mean values, while the squared shaded areas represent the corresponding standard deviation. The flux density was extracted from circular regions of high-resolution images, as is depicted in Fig. \ref{Filaments-regions}.}
    \label{FilamentsFluxSIHighResolution}
\end{figure*}

In Fig. \ref{FilamentsFluxSIHighResolution} in the left and middle panels, the surface brightness profiles for F1 and F2 are presented for high-resolution images together with the mean and standard deviation. The left panel of Fig. \ref{FilamentsFluxSIHighResolution} corresponds to the LOFAR profile at 144 MHz, and the middle panel represents uGMRT at 400 MHz. As was mentioned earlier, the origin of these two filaments is unknown. As opposed to the head-tail, the first and last circular regions for flux density extraction were chosen arbitrarily, each with a beam size of 6\arcsec \ , as is illustrated in Fig. \ref{Filaments-regions}. These regions were drawn based on the $3\sigma$ emission of the images.

The mean surface brightness values and their corresponding standard deviation for F1 and F2 for high-resolution images are shown in Table \ref{tab:FluxScatteringHighLowresolution}. The standard deviation is slightly higher in uGMRT data, particularly at F2.

\begin{table*}[h!]
  \centering
  \captionsetup{labelfont=bf}
  \caption{Mean surface brightness and spectral index, along with their standard deviations for filaments F1 and F2.}
  \label{tab:FluxScatteringHighLowresolution}
  \begin{tabular}{
    |c|c|c|c|
  }
    \toprule
    \multirow{2}{*}{\text{Filaments}} & \multicolumn{1}{c|}{\text{Lofar SB}} & \multicolumn{1}{c|}{\text{uGMRT SB}} & \multicolumn{1}{c|}{\text{Spectral Index}} \\
    & \multicolumn{1}{c|}{\text{[$\text{Jy/arcsec}{^2}$]}} & \multicolumn{1}{c|}{\text{[$\text{Jy/arcsec}{^2}$]}}  & \multicolumn{1}{c|}{} \\
    \midrule
    F1 & Mean: 1.39 $\times \ 10^{-4}$  & Mean: 1.04 $\times \ 10^{-5}$  & Mean: 2.57\\
       & Std. Dev.: 0.32 $\times \ 10^{-4}$ & Std. Dev.: 0.37 $\times \ 10^{-5}$ & Std. Dev.: 0.27\\
    F2 & Mean: 1.36 $\times \ 10^{-4}$ & Mean: 8.49 $\times \ 10^{-6}$ & Mean: 2.78\\
       & Std. Dev.: 0.34 $\times \ 10^{-4}$ & Std. Dev.: 4.47 $\times \ 10^{-6}$ & Std. Dev.: 0.22\\
    \bottomrule
  \end{tabular}
  \tablefoot{These measurements were performed from high-resolution images of LOFAR at 144 MHz and uGMRT at 400 MHz.}
\end{table*}

In the right panel of Fig. \ref{FilamentsFluxSIHighResolution}, we show the low-frequency spectral index trends for F1 and F2 for high-resolution images using the same circular regions as the one used to compute the flux before (see Fig. \ref{Filaments-regions}). In the F1 profile, there is an indication of steepening from regions 4 to 15.

The mean value of the spectral index and the scatter from it between LOFAR and uGMRT, for high-resolution observations, are shown in the third column of Table. \ref{tab:FluxScatteringHighLowresolution}. The standard deviation from the mean value for both F1 and F2 with LOFAR and uGMRT is similar, within $\sim 20 \%$.

\section{Discussion \label{4}}
In this work, we have studied the head-tail radio galaxy and the revived fossil plasma within the galaxy cluster A1775. 
With the higher sensitivity and higher resolution of the new uGMRT data, together with a wider frequency range of low frequencies, we have better constrained the steep spectrum of the outer part of the tail compared to previous studies. We have been able to obtain spectral index and curvature maps of the central region of A1775 at high frequency that have not been obtained before. Using regions drawn along the ridge line of the tail in low-resolution images, we can measure flux densities and obtain surface brightness and spectral index profiles. We analyzed the radio colour-colour diagrams for low-resolution images and compared them with different ageing models. Due to the minor detection of the revived fossil plasma in the uGMRT 650 MHz image, we could only generate spectral index profiles using low-frequency observations, which had not previously been done.

Notably, in the case of the head-tail radio galaxy, we observe an unexpected increase in surface brightness and a steepening of the spectral index in some regions along the tail, which we address here. Additionally, we explain the classification of the filaments as revived fossil plasma based on their spectral properties.

In Section \ref{3.2} in the upper panel of Fig. \ref{SpectralIndexPlots}, we observe a prominent peak of surface brightness profiles at the head of the tail. Additionally, the highest surface brightness does not coincide with the first region, which might imply a complex bending of the jets near the head.
The head of the tail radio galaxy B1339+266B was studied at very high resolution by \cite{Terni_de_Gregory_2017} using the VLA. In this study, the radio core is located south of the brightest spot in both 1.4 GHz and 5 GHz. At 5 GHz, the innermost structure of the source was detected as an off-centre nucleus and two opposite symmetric jets. The northeastern jet is
relatively straight, aligned with the direction of the tail, whereas the southwestern jet shows a prominent bend from the core. This structure was confirmed by the 15 GHz data as well \citep{Terni_de_Gregory_2017}. The twin jets of B1339+266BJ show sharp bends very close to the radio galaxy core. This explains why the starting point (region 0) in the surface brightness profiles (referring to the upper panel of Fig. \ref{SpectralIndexPlots}) does not correspond to the peak surface brightness.

We observe an unexpected steepening of the spectral index in the inner regions of the tail, as is seen in the bottom panel of Fig. \ref{SpectralIndexPlots}.
Considering that these regions are close to filament F1 (refer Fig. \ref{Head-tail-regions}), we suggest that the filament emission contaminates the head-tail radio galaxy. Further support comes from the fact that this steepening is less pronounced in the high-frequency spectral index, and we know that F1 and F2 have steep spectra, hence their contribution to the emission is lower.

In the colour-colour diagram (Fig. \ref{fig:Color-color with model}), the ‘classic’ ageing models assume that particles progressively lose their energy because of radiative losses due to inverse Compton and synchrotron emission. Both the JP and Tribble models can approximately reproduce the data
points of the entire length of the tail, with only some deviations of certain data points, whereas the KP model fails to describe the data points
beyond $\sim$ 450 kpc. Similar trends of data points on colour-colour diagrams, when compared with the same ageing models, have been observed for other head-tail radio galaxies \citep[e.g.,][]{Lusseti_2024, Bruno2024}.

We computed the maximum distance, $d$, that the relativistic plasma can travel by multiplying the radiative age with the velocity in the plane of the sky of the head-tail radio galaxy. We calculated this velocity for our radio galaxy moving through the ICM using the equation
\begin{equation}
    v_{ps}=v_{los} \cdot \sqrt{2}
    \label{VPS}
.\end{equation}
Here, $v_{los} = \Delta z \cdot c$ represents the line-of-sight velocity, $\Delta z$ is the difference between the redshift of the galaxy cluster and
of the radio galaxy, and $c$ is the speed of light. The redshift value for the head-tail radio galaxy B1339+266B (also known as UGC
08669 NED02) was obtained from NED\footnote{\url{https://ned.ipac.caltech.edu}} , and it is z = 0.06942. Together with the redshift of Abell 1775, z = 0.07203, we obtained a velocity of $v_{ps}\simeq1105 \ \text{km/s}$ for the head-tail radio galaxy, which is consistent with typical values for these radio galaxies \citep{Sebastian2017}.

We computed the radiative lifetime of particles emitting at each central frequency using the revised equipartition magnetic field, $B'_{eq}$, and $B_{CMB}$. The radiative lifetime of relativistic electrons undergoing synchrotron and inverse Compton losses is:

\begin{equation}
    t_r \simeq 3.2 \cdot 10^{10} \frac{B_{eq}^{1/2}}{ {B'_{eq}}^{2} + B_{CMB}^2} \frac{1}{\sqrt{\nu (1+z)}} \ \ \text{(yr)} \ .
    \label{TR}
\end{equation}

Here, $B_{eq}$ and $B_{CMB}$ are in units of microgauss and frequency $\nu$ is in megahertz.
Then, by multiplying the Eq. \ref{VPS} and \ref{TR}, we found that the maximum distance that the relativistic plasma can travel is $\sim 230$ kpc. This distance is significantly shorter than the total length of the tail at all three frequencies, which exceeds 600 kpc, suggesting that the emitting electrons are re-accelerated along the tail. The fact that the data points in the colour-colour diagram are broadly in line with a JP and Tribble model may be justified in a scenario where the time-period of re-acceleration is not much longer than the re-acceleration time of relativistic electrons.

With a deeper analysis of spectral index profiles for the F1 and F2 filaments (shown in the previous Section), we provide additional evidence supporting the classification of F1 and F2 as revived fossil plasma, based on their distorted, irregular morphology and ultra-steep spectrum, $\alpha > 2$, which lacks any clear optical counterpart(s). From the comparison of radio and X-ray data in the \citet{Botteon_2021} study, it was pointed out that these filaments likely trace regions where the thermal gas is compressed, resulting in the re-energization of the plasma. The double giant elliptical system at the cluster's centre is a strong candidate for the origin of these filaments, which may have previously injected relativistic plasma into the ICM, which was subsequently revived by gas dynamics triggered by core motions. Based on the spectral profile seen in the right panel of Fig. \ref{FilamentsFluxSIHighResolution}, one can speculate a steepening of F1 from the region near the elliptical galaxies toward region 17, possibly indicative of a relation between the revived plasma and the double giant elliptical system.

\subsection{Underestimated magnetic field}
The estimation of the magnetic field along the tail in Section \ref{3.2.2} was computed by using the revised equipartition formula, a method that relies on several assumptions, such as the value $k=1$, which is usually assumed in the literature for galaxy clusters and head-tail radio galaxy studies (see for example \citealt{Sebastian2017}), as well as the study concerning the head-tail radio galaxy NGC 4869 in the Coma cluster by \cite{Lal_2020}.

As electrons age along the tail, they lose most of their energy, unlike CR protons. This difference likely causes the increase in the value of $k$  along the tail (though the exact trend of this increase remains unknown). Together with the assumptions we made about the depth and the lower cut-off, $\gamma_{min}$, we suggest that the values of the magnetic field along the tail are likely to be underestimated. The magnetic field profile along the tail of head-tail galaxies has not been investigated much in the literature. Studies of the magnetic field derived from equipartition in various tails have shown values consistent with our findings. The values we obtained here are comparable to the ones in the study of NGC 4869 head-tail by \cite{Lal_2020} and the sample of head-tail radio galaxies by \cite{Sebastian2017}, obtained following the same methodology, except for the regions within the head of the tail where the jets exhibit complex bending, which could not be resolved with our observation.

\subsection{Pressure balance}
We computed the minimum pressure for the head-tail radio galaxy using the standard relation between pressure and energy density for relativistic particles:
\begin{equation}
    P_{eq}=U_{min}/3 \ .
\end{equation}

We aim to compare this projected internal pressure of the head-tail radio galaxy with the external ICM pressure profile of central A1775 from \cite{Hu_2021}. This comparison helps us to assess if there is pressure equilibrium and infer where the head-tail radio galaxy is spatially located within the cluster.

\begin{figure}[h!]
    \centering
    \captionsetup{labelfont=bf}
    \includegraphics[width=0.5\textwidth]{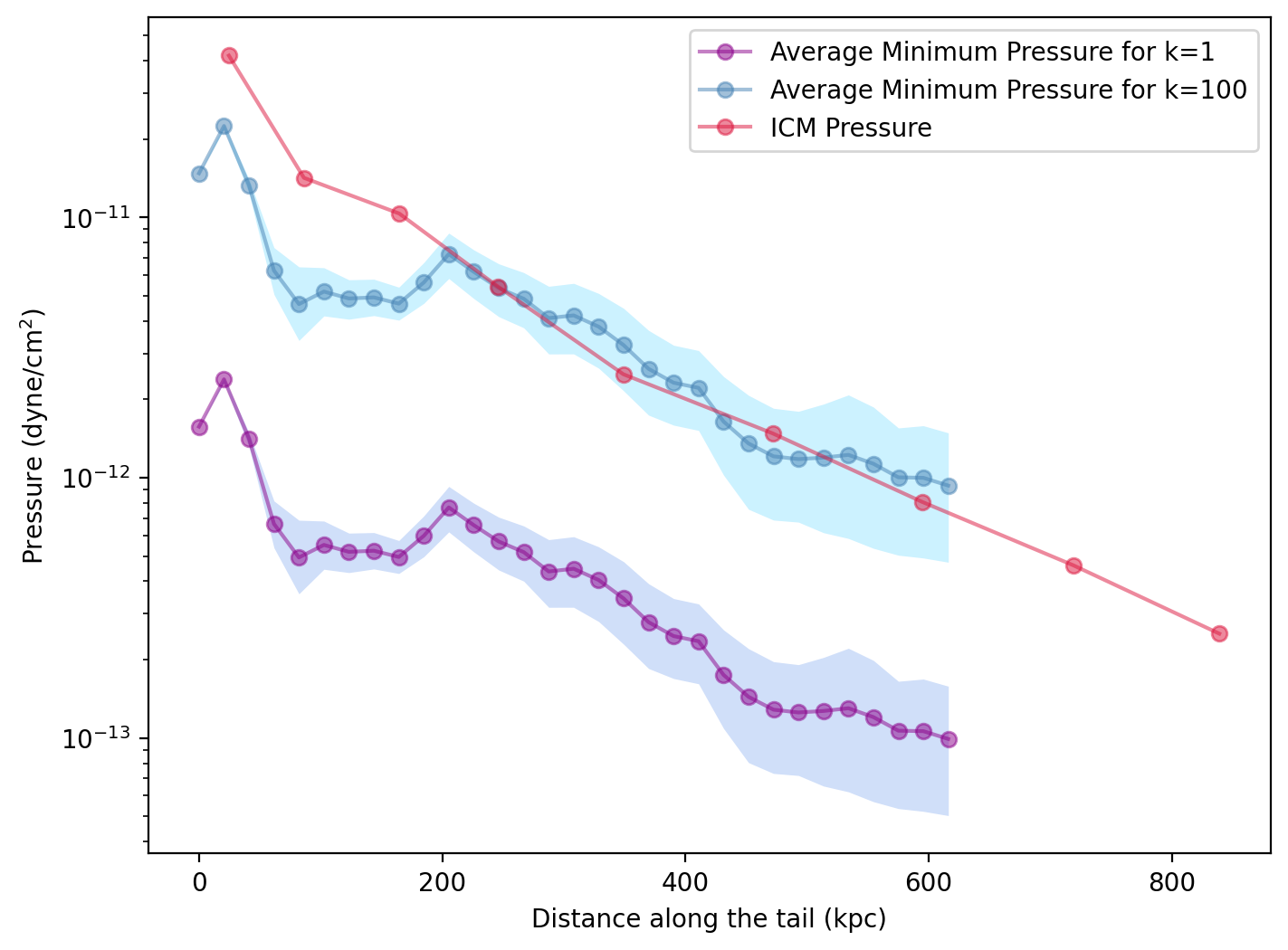}
       \caption{Three different profiles of pressure in dyne per square centimetre over the distance along the tail in kiloparsecs. The violet profile is the average profile of minimum pressure assuming $k=1$ between low-resolution images at three frequencies, LOFAR at 144 MHz and uGMRT at 400 MHz and 650 MHz. The shaded area is the minimum and maximum of all three profiles. The blue profile is similar to the violet one but assumes $k=100$. The red profile is the external ICM profile inferred from XMM-Newton observations.}
    \label{ICM_pressure}
\end{figure}

We used the average minimum-pressure obtained profiles between all three frequencies in dyne per square centimetre unit.
We computed the profile of the external ICM pressure using the averaged projected gas temperature and the corresponding electron number density distribution profiles obtained by \cite{Hu_2021} using XMM-Newton data, measured from the centre of the cluster. In Fig. \ref{ICM_pressure}, these two profiles, shown in red and violet, are plotted along the projected distance along the tail.

This figure assumes that the head is at the centre of the cluster, that the tail is in the plane of the sky, and that there appears to be no pressure balance. If the tail has a different orientation, the distances should be increased by a factor of $\sqrt{2}$, moving outwards from the observed points. Since the minimum pressure was derived from the minimum energy condition, several assumptions were made regarding $k$, the extent of the source along the line of sight, $d$, and the volume filling factor, $\Phi$, which might lead to the observed pressure imbalance. 
For example, the composition of the jets is believed to change along the tail, with protons becoming energetically more dominant at larger distances \citep[e.g.,][]{Croston2014, Croston2018}. If in our calculation we assume $k=100$, the two pressures are nearly balanced, starting from region 10 (at $\sim 200 kpc$), where the tail begins to slightly expand in width until the end of the tail (see the blue and red profiles in Fig. \ref{ICM_pressure}). We note, however, that any constant value of $k$ would likely be an oversimplification, as $k$ should evolve along the jets. The unknown 3D location of the head-tail within the galaxy cluster is another factor that can contribute to the observed pressure imbalance.

\section{Conclusions \label{5}}
In this work, we have used low-frequency data to perform a deeper investigation of the old relativistic plasma in the central region of the galaxy cluster Abell 1775. This study focuses on the extension of the head-tail radio galaxy and the revived fossil plasma filaments F1 and F2, using LOFAR data at 144 MHz and new deep uGMRT data at 400 and 650 MHz.

From all the spectral measurements and analysis conducted, the main conclusions in this work are summarized as follows:

\begin{enumerate}
    \item We speculate that the regions corresponding to the head of the tail, which show an unexpected increase in surface brightness, might come from the complex bending of the jets near the head. The head of the tail B1339+266B was previously studied at very high resolution by \cite{Terni_de_Gregory_2017} using the JVLA. It was concluded that the twin jets show sharp bends very close to the radio galaxy core.
    \item In the colour-colour diagram comparison with the ageing models of low-resolution images, we find that the JP and Tribble models can reproduce the trend of the data points across the entire length of the head-tail, with only minor deviations observed at some data points. Our new deep low-frequency observations support the scenario of ongoing particle re-acceleration in the outer region of the tail, as is suggested by \cite{Botteon_2021}, as the extent of the tail is greater (by about a factor of 3) than the distance that the relativistic plasma can travel. The re-acceleration is likely to occur at the position of the cold front, where the tail appears to deviate direction and changes its brightness structure at all three frequencies of our observations.
    \item We conclude that the estimated equipartition magnetic field from the revised formula is decreasing along the tail and that it is comparable with typical values found in other studies of different head-tail radio galaxies \citep[e.g.,][]{Sebastian2017,Lal_2020}, except at the head of the tail, where the derived magnetic field strength is likely higher due to the complex bending of the jets close to the host galaxy.
    \item The combination of different factors, such as the assumptions made in the minimum energy equation, project effects, and the unknown 3D location of the head-tail within the galaxy cluster, contributes to the observed imbalance between the internal and external pressures. However, when $k$ is increased to 100, a near balance is observed between the pressures of the head-tail and the ICM, except in the innermost regions of the tail.
    \item We conclude that both filaments F1 and F2 have an ultra-steep spectrum with $\alpha>2$, based on only the low-frequency spectral index profiles. The high-frequency spectral index profiles would be even steeper. Their origin is consistent with them being fossil plasma that has been re-energized by processes in the ICM. We speculate that the steepening of the spectral profile of filament F1 may indicate that its origin is related to one of the giant elliptical galaxies, whereas the spectral profile of F2 does not show any clear trend.
\end{enumerate}

The results of this paper show the power of combining deep radio observations at low frequencies in the study of steep spectrum synchrotron emission sources in galaxy clusters. High-sensitivity and high-angular-resolution observations at low frequencies are crucial for studying in detail the resolved properties of tails and the structure of revived fossil plasma in the ICM.

As the jets of the extended head-tail radio galaxy in A1775 show complex bends, likely affecting the first data point of our profiles, we anticipate that employing LOFAR international baselines, which would allow us to reach resolutions of $\sim0.3"$, will enable us to resolve the structure of the jets on the scales probed by \cite{Terni_de_Gregory_2017}. This resolution will allow us to study the jets starting from the black hole and extending beyond the optical galaxy into the region we refer to as the ‘inner tail’.

\begin{acknowledgements}
We thank the referee for constructive comments that improved the manuscript. We thank Gianfranco Brunetti for the useful discussion.
LOFAR \citep{vanHaarlem2013} is the LOw Frequency ARray designed and constructed by ASTRON. It has observing, data processing, and data storage facilities in several countries, which are owned by various parties (each with their own funding sources), and are collectively operated by the ILT foundation under a joint scientific policy. The ILT resources have benefitted from the following recent major funding sources: CNRS-INSU, Observatoire de Paris and Universit\'{e} d'Orl\'{e}ans, France; BMBF, MIWF-NRW, MPG, Germany; Science Foundation Ireland (SFI), Department of Business, Enterprise and Innovation (DBEI), Ireland; NWO, The Netherlands; The Science and Technology Facilities Council, UK; Ministry of Science and Higher Education, Poland; Istituto Nazionale di Astrofisica (INAF), Italy. This research made use of the Dutch national e-infrastructure with support of the SURF Cooperative (e-infra 180169) and the LOFAR e-infra group, and of the LOFAR-IT computing infrastructure supported and operated by INAF, and by the Physics Dept.~of Turin University (under the agreement with Consorzio Interuniversitario per la Fisica Spaziale) at the C3S Supercomputing Centre, Italy. The J\"{u}lich LOFAR Long Term Archive and the German LOFAR network are both coordinated and operated by the J\"{u}lich Supercomputing Centre (JSC), and computing resources on the supercomputer JUWELS at JSC were provided by the Gauss Centre for Supercomputing e.V. (grant CHTB00) through the John von Neumann Institute for Computing (NIC). This research made use of the University of Hertfordshire high-performance computing facility and the LOFAR-UK computing facility located at the University of Hertfordshire and supported by STFC [ST/P000096/1]. We thank the staff of the GMRT for making these observations possible. GMRT is run by the National Centre for Radio Astrophysics of the Tata Institute of Fundamental Research. RJvW acknowledges support from the ERC Starting Grant ClusterWeb 804208. MB acknowledges funding by the Deutsche Forschungsgemeinschaft (DFG) under Germany's Excellence Strategy – EXC 2121 Quantum Universe – 390833306 and the DFG Research Group "Relativistic Jets".
Basic research in radio astronomy at the Naval Research Laboratory, which is supported by 6.1 Base funding.
\end{acknowledgements}

%
%

\bibliographystyle{aa}
\bibliography{references}

\begin{appendix}
\section{Spectral index and curvature uncertainty maps \label{A}}
\begin{figure}[h!]
    \centering
    \captionsetup{labelfont=bf}
    
    \begin{subfigure}{0.225\textwidth}
        \centering
        \includegraphics[width=1.0\textwidth]{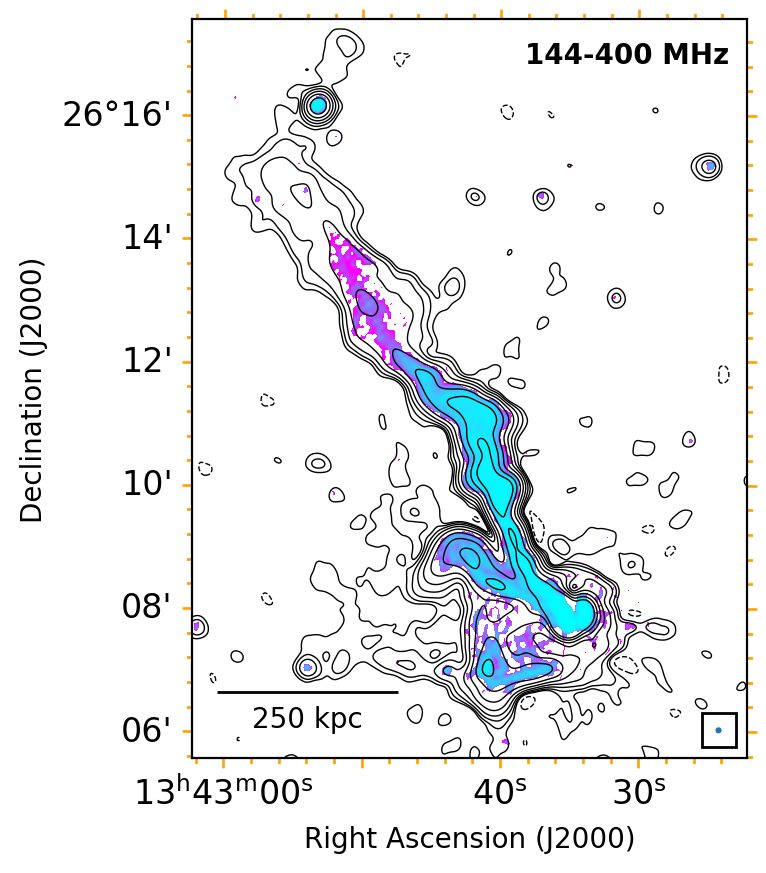}
    \end{subfigure}
    \hfill
    \begin{subfigure}{0.252\textwidth}
        \centering
        \includegraphics[width=1.0\textwidth]{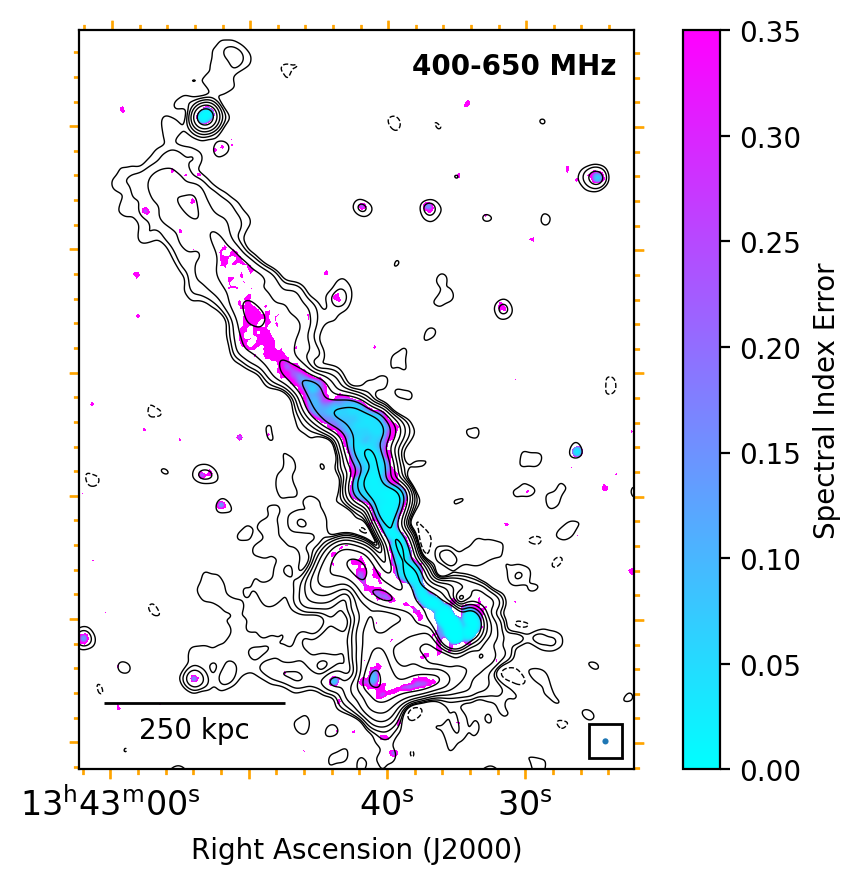}
    \end{subfigure}
    
    \begin{subfigure}{0.225\textwidth}
        \centering
        \includegraphics[width=1.0\textwidth]{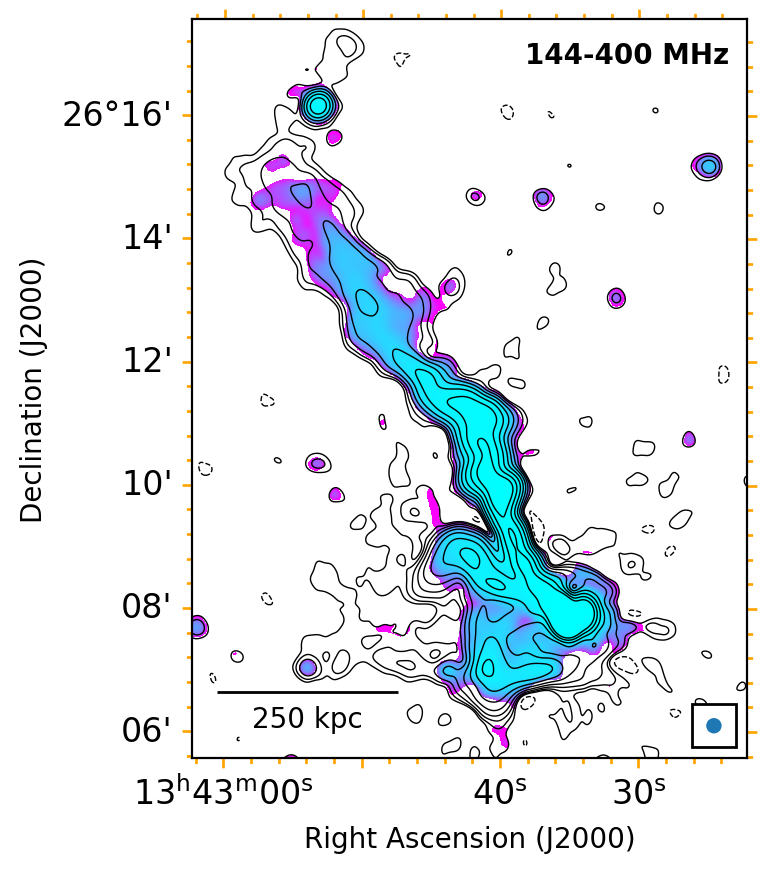}
    \end{subfigure}
    \hfill
    \begin{subfigure}{0.252\textwidth}
        \centering
        \includegraphics[width=1.0\textwidth]{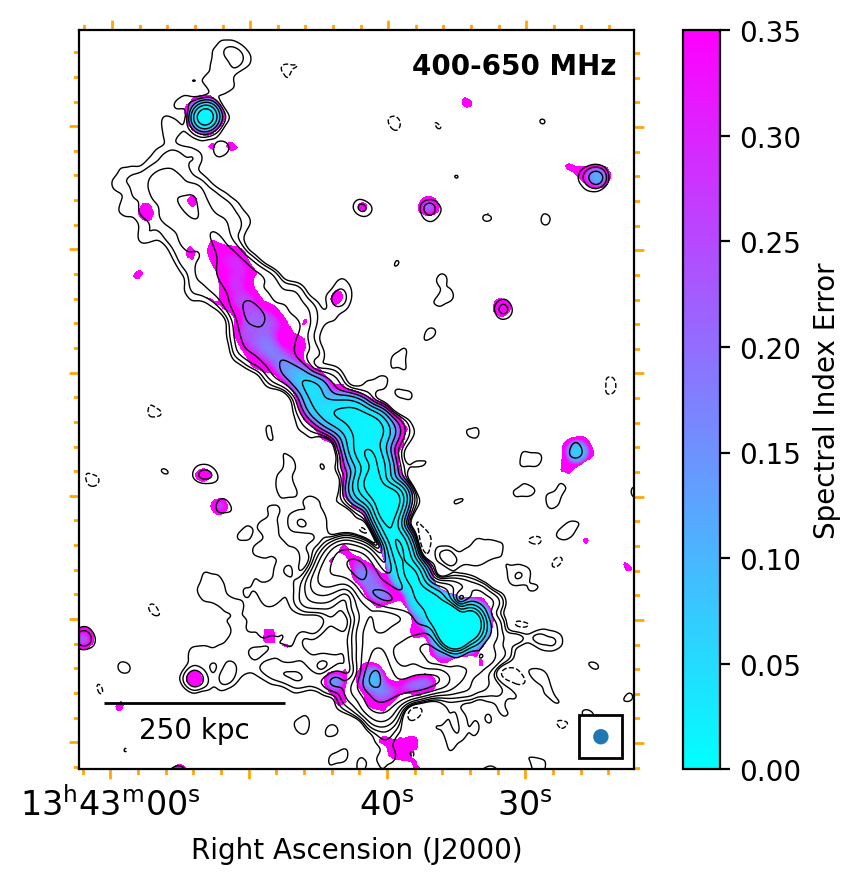}
    \end{subfigure}
    
    \caption{Spectral index uncertainty maps corresponding to Fig.\ref{SpectralIndexMaps}.}
    \label{Spectral index error maps}
\end{figure}

\begin{figure}[h!]
    \centering
    \captionsetup{labelfont=bf}
    \begin{subfigure}{0.2455\textwidth}
        \centering
        \includegraphics[width=1.0\textwidth]{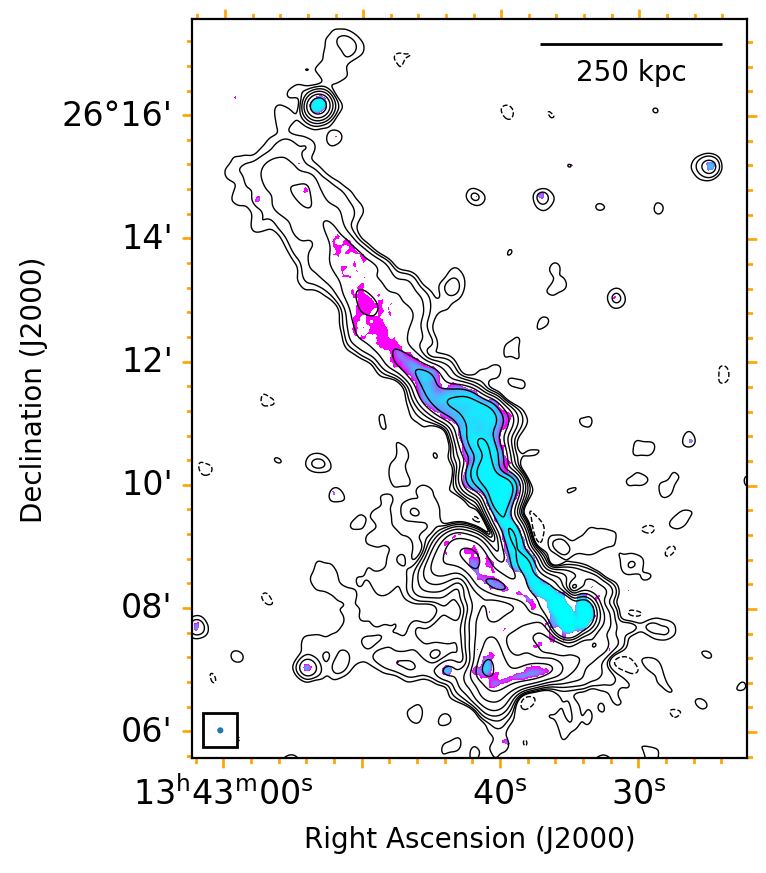}
    \end{subfigure}%
    \begin{subfigure}{0.2555\textwidth}
        \centering
        \includegraphics[width=1.0\textwidth]{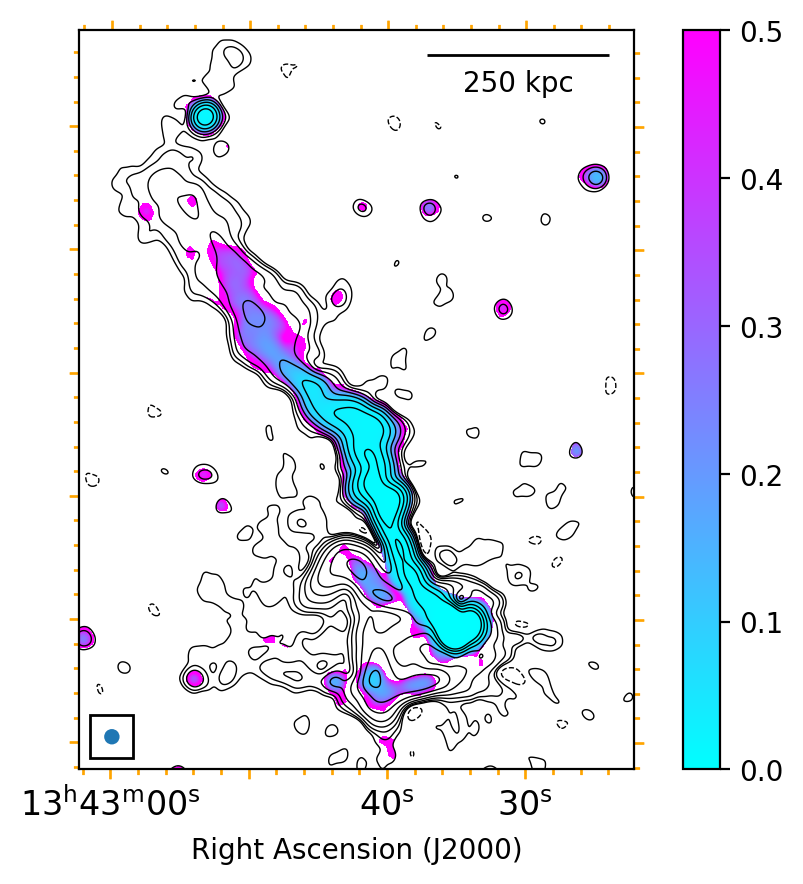}
    \end{subfigure}%
    \caption{SC uncertainty maps corresponding to Fig.\ref{CurvatureMap}.}
    \label{CurvatureErrorMaps}
\end{figure}

In Figures \ref{Spectral index error maps} and \ref{CurvatureErrorMaps}, we show the spectral index uncertainty maps and the SC uncertainty maps corresponding to both high- and low-resolution images. We note that the uncertainty values across the central region of A1775 are generally small, with a slight increase observed in the outer tail and revived fossil plasma.

\end{appendix}

\end{document}